%% file: main.tex
\documentclass{article}

\usepackage{my_style}
\usepackage{setspace}

\usepackage[T1]{fontenc}
\usepackage[utf8]{inputenc}
\usepackage{lmodern}

\makeatletter
\renewcommand\@fnsymbol[1]{\ensuremath{%
  \ifcase#1\or\dagger\or\ddagger\or\mathsection\or\mathparagraph\or
  \|\or\*\or\dagger\dagger\or\ddagger\ddagger\else\@ctrerr\fi}}
\makeatother

\title{Taxing Artificial Intelligence}
\author{Juliette Faivre and Sarah H. Cen \\
Carnegie Mellon University
}
\date{}

\begin{document}
\maketitle

\begin{abstract}
    While AI promises major benefits, its development and deployment can shift costs onto others, including environmental pressures on local communities, labor and creative displacement, and systemic risks from rapid frontier development.
    Taxation is an integral part of policy design, and recent academic, industry, and policy debates have begun to consider whether tax instruments can help address these harms. 
    In this paper, we explore the viability of AI taxation. 
    More broadly, AI taxation should not be understood only as Pigouvian correction. 
    In the AI context, taxation can also correct harmful activity, redistribute unevenly borne costs and gains, and fund regulatory capacity.
    We discuss the main externalities associated with AI and survey possible tax instruments, including corporate income and rent-based taxes, consumption taxes on AI-related services, and excise instruments tied to specific AI activities. 
    We further assess the benefits and pitfalls of these instruments, including feasibility, measurement problems, incidence, leakage, and innovation costs. 
    Because AI externalities differ in nuanced ways, tax policy must be carefully designed and matched to the specific harms and policy objectives. 
\end{abstract}
\vspace{1.0em}
\setcounter{tocdepth}{3}
\tableofcontents
\clearpage

\input{sections/intro_new}

\input{sections/case_studies}

\input{sections/background}

\input{sections/ai_externalities}
\input{sections/case_for_ai_taxes}

\input{sections/surveying_options}
\input{sections/pitfalls_open_questions}
\input{sections/conclusion}
\input{sections/acknowledgements}

\bibliography{bib.bib}

\end{document}

%% file: sections/intro_new.tex
\section{Introduction}

Taxation is an unpopular concept. 
It creates frictions in a free market, eats into earnings, and inflates the cost of transactions, often for benefits that are invisible to the taxpayer. 
Yet there are cases where taxation is justifiable, and artificial intelligence (AI) may be one of them.

The growing momentum behind AI taxation makes this question increasingly urgent. 
In just a few years, AI taxation has moved from a largely theoretical debate around ``robot taxes'' (see \Cref{ssec:related_work}) into concrete discussions among federal lawmakers, state governments, and AI industry leaders. 
At the federal level, Senator Bernie Sanders introduced the American A.I. Sovereign Wealth Fund Act~\cite{sanders2026publicownai, sanders2026american_ai_sovereign_wealth_fund}, which would impose a one-time stock-based tax on large AI companies to seed a public sovereign wealth fund if passed. 
Similarly, Senator Mark Kelly's \textit{AI for America} roadmap proposes an AI Horizon Fund~\cite{kelly2025ai_for_america}, financed by contributions from technology firms, to support workers, infrastructure, and environmental mitigation. Senator Elizabeth Warren has called for an AI tax~\cite{Ft2026EquityStakeAI}, with attention to data center energy usage. 
At the state level, Virginia lawmakers approved what appears to be the first U.S. tax tied directly to data center electricity use~\cite{virginia2026_data_center_electricity_tax}, while Ohio and Utah have likewise moved to limit or condition data center incentives, reflecting concerns around their social impacts~\cite{cox2026_higher_bar_data_center_development, dewine2026_pause_data_center_tax_exemption}. Industry leaders have also begun to frame AI-driven wealth and displacement as tax questions. 
Dario Amodei has emphasized the role of taxation in redistribution under possible AI-driven displacement~\cite{amodei2024adolescence_technology, amodei2026_policy_ai_exponential}, while Sam Altman's American Equity Fund proposal, revived in a recent OpenAI policy paper~\cite{openai_industrial_policy_intelligence_age}, would distribute the gains from AI capital ownership. These developments suggest that AI taxation is no longer a speculative policy idea.

Underlying this momentum is a growing recognition that AI imposes costs on parties who do not share in its gains. 
AI has emerged as a foundational technology, deeply integrated into modern social and economic systems~\cite{maslej2025artificial}. 
Its promise is substantial, particularly in fields such as scientific discovery, labour productivity, and healthcare~\cite{wang2023scientific, brynjolfsson2025generative, alowais2023revolutionizing}. 
However, AI also generates harms that extend beyond faulty outputs or individual misuse, encompassing broader societal externalities associated with its development and deployment~\cite{international_ai_safety_report2025, bommasani2021opportunities}. 
These include pressures on the environment and local communities, the labor market, creative industries, and systemic risks. Moreover, 
\emph{these same features that make AI valuable, such as its speed, scale, and increasing autonomy, also accelerate the diffusion of its harms} ~\cite{bengio_hinton_yao2024extreme_ai_risks}.

In response to growing and diffuse AI harms,
policymakers have begun to propose and adopt an assortment of regulatory tools. These include risk-management frameworks~\cite{nistrmfprinciplesai2023artificial, kaminski2023regulating}, transparency obligations~\cite{california2024sb942, bommasani2024foundation, bommasani2023foundation}, conformity assessments~\cite{conformityneuman2021european, mokander2022conformity, floridi2022capai}, incident reporting~\cite{california2025sb53, gailmard2025known, dixon2025ai, mcgregor2021preventing}, auditing~\cite{auditingnyscomptroller2025locallaw144, raji2020closing, kroll2015accountable}, regulatory sandboxes~\cite{sandboxtexas2025hb149analysis, nabil2024artificial, buocz2023regulatory}, and principles for trustworthy AI~\cite{fjeld2020principled, jobin2019global}. 
Yet these efforts remain challenging to implement, primary among the reasons being enforceability~\cite{brundage2026frontier}. 
For many proposals, the regulatory infrastructure remains nascent~\cite{guha2024ai}, and many approaches require sophisticated audit, measurement, and evaluation technologies that are themselves nascent~\cite{mokander2023auditing, ojewale2025towards}. 
Developing these tools requires technical expertise, but governments often struggle to compete with the private sector for AI talent~\cite{gao2026artificialintelligence}, and much of this capacity is underfunded. 
All of these factors exacerbate the delay in oversight, which already lags behind state-of-the-art technologies~\cite{international_ai_safety_report2025}.
\\

In this paper, we investigate the role that taxation could play in AI governance. 
In doing so, we address five questions: (i) what externalities has AI introduced; (ii) what is the purpose of taxing AI; (iii) how does taxation fit into the broader set of regulatory approaches; (iv) how should AI taxes be designed; and (v) what are the pitfalls of taxation? 

A central finding of this paper is that there are three strong reasons to tax AI.
The first is to \textbf{correct harmful activity by pricing it and forcing actors who generate a social harm to internalize it}.
Doing so reshapes incentives and thus encourages these actors to change their behavior.
Taxation is a particularly effective corrective tool, but this also implies that taxation must be done carefully or otherwise risk causing unintended consequences (e.g., gaming behaviors or tax burdens on the wrong actors).
The second is to \textbf{redistribute AI's costs and gains, which fall unevenly across society}. 
Actors who generate the social harms of AI are generally not the ones who experience them. 
Residents facing higher utility bills near data centers, creators whose work is used to train AI models, workers replaced by AI systems, and individuals whose personal and private data are released are usually not the ones who profit from the AI activities that bring about these outcomes. 
Tax revenue can be directed back toward these groups, or toward broader needs such as education, worker transition programs, infrastructure upgrades, and welfare.
The third is to \textbf{fund regulatory capacity}, especially AI oversight. 
Many AI proposals and existing regulations rely on monitoring, auditing, expertise, and enforcement that require funding sources that are elusive. 
A tax on the AI sector \textbf{creates a steady source of revenue for AI regulatory capacity}.

\input{sections/figure_summary}

Taxation is also a familiar mechanism, and that familiarity comes with several practical advantages. \textbf{Firms already know how to file, pay, and respond to tax obligations, and governments already know how to assess, collect, audit, and enforce them} ~\cite{oecd2025tax_administration}. An AI tax can therefore build on existing infrastructure rather than waiting for new oversight programs to be designed from scratch. In practice, this means regulation through taxation can be {implemented quickly}. The tax system also includes independent public auditing, giving enforcement a structure that many newer regulatory approaches often lack.
Because tax compliance is already part of ordinary business operations, even small firms can often meet new obligations without hiring an entirely new compliance team, as AI-specific regulatory frameworks may require~\cite{deluca2007irs_research_conference}. Taxation can also shape behavior through prices on measurable activities {without prescribing technical requirements for AI systems or resolving contested AI definitions}.

In the U.S., there is one additional and significant advantage: Congress can pass major fiscal legislation, including tax bills, with a simple majority rather than the supermajority often required to overcome a filibuster~\cite{crs2024budget_reconciliation}. 
This pathway---known as \textbf{budget reconciliation}---makes federal tax legislation a more viable path than many non-fiscal regulatory proposals when Congress is otherwise gridlocked.

We further examine AI externalities and possible tax design in \Cref{sec:ai_externalities,sec:designing_tax}. 
In \Cref{sec:ai_externalities}, we review the main social harms of AI---including rising water and electricity prices, labor displacement, environmental costs, misinformation and hallucinations, bias and discrimination, privacy risks, cybersecurity and system vulnerabilities, and catastrophic risk---and discuss how each lines up with the definition of an externality. 
In \Cref{sec:designing_tax}, we review the main considerations in designing an AI tax, including the choice of policy objective, tax base, taxpayer, instrument, and rate. 
We then consider different AI tax instruments, including excise taxes, payroll taxes, consumption taxes, and corporate or rent-based taxes.

At the same time, taxation is no panacea. It should be understood as complementary to other AI governance tools, not as a substitute for them. 
Moreover, it carries real limitations, such as measurement difficulties, incidence effects, jurisdictional leakage, political capture, and potential costs to innovation and competitiveness. 
Some harms, particularly rights-based harms such as privacy loss and discrimination, may be poor candidates for taxation altogether. We take these concerns seriously and return to them in \Cref{sec:pitfalls}.

The remainder of the paper proceeds as follows. 
\Cref{sec:case_studies} introduces case studies that illustrate possible targets for AI taxation. 
\Cref{sec:background} provides background on the purposes and design of tax instruments, including revenue-raising and corrective taxation. 
\Cref{sec:ai_externalities} maps the main externalities associated with AI, including rising water and electricity prices, labor displacement, and other potential externalities and risks. 
\Cref{sec:case_for_AI} develops the practical case for AI taxation, arguing in favor of the existing and established enforcement infrastructure, the relative independence of tax administration, the availability of professional accounting expertise, and the flexibility of tax design. \Cref{sec:designing_tax} surveys possible AI tax instruments that may be considered, including corporate income taxes, consumption taxes, and excise taxes.
We also review important considerations when taxing, including that selecting the right AI actor and activity along the AI supply chain is critical but difficult because ``AI-ness'' is difficult to define.
Finally, \Cref{sec:pitfalls} considers the main pitfalls and limitations of AI taxation, including innovation and global competitiveness costs, various distortions caused by taxation, and measurement challenges.

\subsection{Related work}\label{ssec:related_work}

While often overlapping, the current discussion on AI taxation grew partly out of earlier literature on robot and automation taxation. Robot taxation literature centered around two related questions: (i) whether current tax systems favor capital-intensive automation over human labor~\cite{kovacev2020taxing}, and (ii) how fiscal policy can respond to labor displacement and a shrinking labor-based revenue. Deliberations included revising the tax treatment of automation and capital~\cite{abbott_bogenschneider2018robots_taxes, mazur2018taxing, ooi_goh2022taxation_automation_ai, dimitropoulou2024robot}, restoring tax neutrality between workers and machines~\cite{abbott_bogenschneider2018robots_taxes, dimitropoulou2024robot}, and considering whether robots could be treated as taxable persons~\cite{oberson2017taxing}. Economic models also examined how fiscal policy instruments on robot adoption affect wages, inequality, and welfare~\cite{guerreiro2022should,thuemmel2023optimal}. On a more practical level, empirical work evaluated the first ``robot tax''~\cite{robottaxgonzalez2022they} policy in South Korea, an automation-investment tax credit, and suggested its potential cost-effectiveness for employment creation~\cite{robottaxkang2024welfare}. Other scholarship remained skeptical of taxation, arguing that robot and automation taxes may be difficult to define, administer, and justify~\cite{chand2020taxing}.

Most recent work turns from industrial automation toward AI-specific public finance. It inherits concerns about labor displacement, tax neutrality, and the erosion of labor-based revenues. However, literature by~\citet{harpaz2026taxingai, korinek2026publicfinance} broadens the debate to how AI can affect revenue raising and redistribution as economic value shifts towards capital. In~\citet{korinek2026publicfinance}, the considered responses include taxes on consumption, capital and rents, compute, tokens, and AI services, alongside sovereign wealth funds and windfall clauses, while~\citet{bearer2025sharing} argue for equity taxes that create public ownership in generative AI firms. Recent work by~\citet{falk2026ai} argues that AI layoffs create a demand externality, motivating a Pigouvian automation tax because firms do not internalize the broader loss of consumer demand from displaced workers. This diversity of tax bases reflects AI's particular nature---intangible, mobile, and embedded across different services---which adds complexity compared to physical machines. This scholarship focuses on tax-base adaptation and the distribution of AI-generated gains, and has not systematically matched fiscal responses to (i) AI's diversified harms and (ii) the policy goals attached to each.
This paper seeks to build on and complement the literature by examining a collection of AI harms and policy objectives that might justify case-specific taxation. 
We identify distinct AI externalities, policy needs, and the corresponding tax design considerations.

%% file: sections/figure_summary.tex
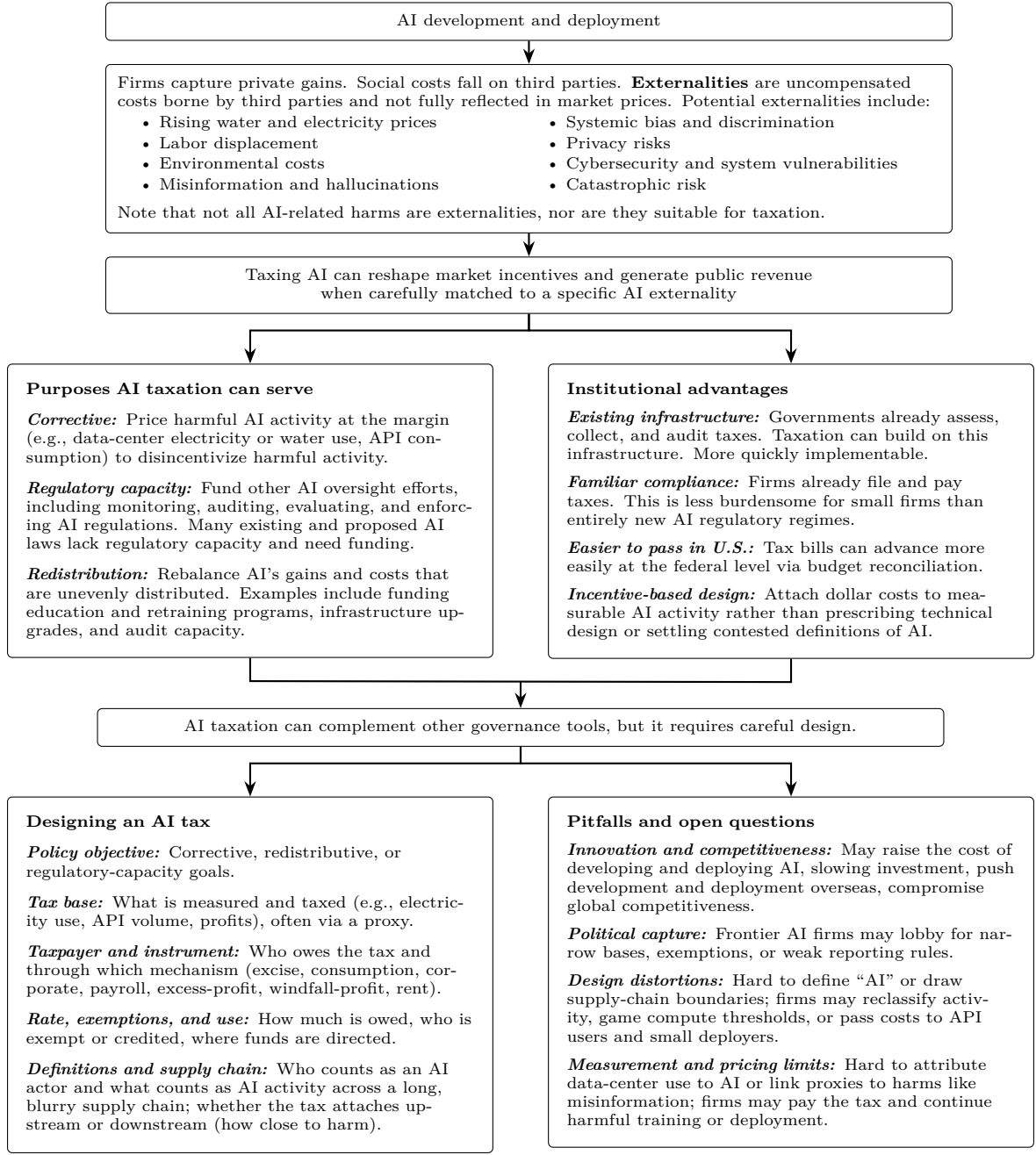
\begin{figure}[htbp]
\centering
\begin{adjustbox}{max width=\textwidth, max totalheight=0.88\textheight, center}
\begin{tikzpicture}[
  font=\scriptsize,
  >=Stealth,
  widebox/.style={
    rectangle,
    rounded corners=2pt,
    draw=black,
    align=center,
    text width=12.4cm,
    inner sep=5pt
  },
  listbox/.style={
    rectangle,
    rounded corners=2pt,
    draw=black,
    align=left,
    text width=12.4cm,
    inner sep=5pt
  },
  channel/.style={
    rectangle,
    rounded corners=2pt,
    draw=black,
    align=left,
    text width=5.8cm,
    inner sep=5pt
  },
  purposesbox/.style={
    rectangle,
    rounded corners=2pt,
    draw=black,
    align=left,
    text width=6.75cm,
    inner sep=8pt
  },
  instbox/.style={
    rectangle,
    rounded corners=2pt,
    draw=black,
    align=left,
    text width=6.75cm,
    inner sep=8pt
  },
  arrow/.style={->, thick}
]

\node[widebox] (ai) {AI development and deployment};

\node[listbox, below=0.35cm of ai] (externalities) {
Firms capture private gains. Social costs fall on third parties.
\textbf{Externalities} are uncompensated costs borne by third parties and not fully reflected in market prices.
{Potential externalities include:}\\[0.15em]
\hspace{1.15em}%
\begin{minipage}[t]{5.55cm}
\textbullet\ Rising water and electricity prices\\[0.1em]
\textbullet\ Labor displacement\\[0.1em]
\textbullet\ Environmental costs\\[0.1em]
\textbullet\ Misinformation and hallucinations
\end{minipage}\hfill
\begin{minipage}[t]{5.95cm}
\textbullet\ Systemic bias and discrimination\\[0.1em]
\textbullet\ Privacy risks\\[0.1em]
\textbullet\ Cybersecurity and system vulnerabilities\\[0.1em]
\textbullet\ Catastrophic risk
\end{minipage}\\[0.65em]
{Note that not all AI-related harms are externalities, nor are they suitable for taxation.}
};

\node[widebox, below=0.35cm of externalities] (tax) {
Taxing AI can reshape market incentives and generate public revenue\\
when carefully matched to a specific AI externality
};

\node[purposesbox, below=0.75cm of tax, xshift=-4.2cm] (purposes) {%
\textbf{Purposes AI taxation can serve}\\[0.65em]
\textbf{\emph{Corrective:}} Price harmful AI activity at the margin (e.g., data-center electricity or water use, API consumption) to disincentivize harmful activity.\\[0.65em]
\textbf{\emph{Regulatory capacity:}} Fund other AI oversight efforts, including monitoring, auditing, evaluating, and enforcing AI regulations. Many existing and proposed AI laws lack regulatory capacity and need funding.\\[0.65em]
\textbf{\emph{Redistribution:}} Rebalance AI's gains and costs that are unevenly distributed. Examples include funding education and retraining programs, infrastructure upgrades, and audit capacity.
};

\node[instbox, below=0.75cm of tax, xshift=3.95cm] (institutional) {%
\textbf{{Institutional advantages}}\\[0.5em]
\textbf{\emph{Existing infrastructure:}} Governments already assess, collect, and audit taxes. Taxation can build on this infrastructure. More quickly implementable.\\[0.5em]
\textbf{\emph{Familiar compliance:}} Firms already file and pay taxes. This is less burdensome for small firms than entirely new AI regulatory regimes.\\[0.5em]
\textbf{\emph{Easier to pass in U.S.:}} Tax bills can advance more easily at the federal level via budget reconciliation.\\[0.5em]
\textbf{\emph{Incentive-based design:}} Attach dollar costs to measurable AI activity rather than prescribing technical design or settling contested definitions of AI.
};

\coordinate (midchannels) at ($(purposes.south)!0.5!(institutional.south)$);

\node[widebox, below=0.75cm of midchannels] (welfare) {
AI taxation can complement other governance tools, but it requires careful design.
};

\node[purposesbox, anchor=north] (design) at ([yshift=-0.75cm]welfare.south -| purposes.center) {%
\textbf{Designing an AI tax}\\[0.65em]
\textbf{\emph{Policy objective:}} Corrective, redistributive, or regulatory-capacity goals.\\[0.65em]
\textbf{\emph{Tax base:}} What is measured and taxed (e.g., electricity use, API volume, profits), often via a proxy.\\[0.65em]
\textbf{\emph{Taxpayer and instrument:}} Who owes the tax and through which mechanism (excise, consumption, corporate, payroll, excess-profit, windfall-profit, rent).\\[0.65em]
\textbf{\emph{Rate, exemptions, and use:}} How much is owed, who is exempt or credited, where funds are directed.\\[0.65em]
\textbf{\emph{Definitions and supply chain:}} Who counts as an AI actor and what counts as AI activity across a long, blurry supply chain; whether the tax attaches upstream or downstream (how close to harm).
};

\node[instbox, anchor=north] (pitfalls) at ([yshift=-0.75cm]welfare.south -| institutional.center) {%
\textbf{Pitfalls and open questions}\\[0.5em]
\textbf{\emph{Innovation and competitiveness:}} May raise the cost of developing and deploying AI, slowing investment, push development and deployment overseas, compromise global competitiveness.\\[0.5em]
\textbf{\emph{Political capture:}} Frontier AI firms may lobby for narrow bases, exemptions, or weak reporting rules.\\[0.5em]
\textbf{\emph{Design distortions:}} Hard to define ``AI'' or draw supply-chain boundaries; firms may reclassify activity, game compute thresholds, or pass costs to API users and small deployers.\\[0.5em]
\textbf{\emph{Measurement and pricing limits:}} Hard to attribute data-center use to AI or link proxies to harms like misinformation; firms may pay the tax and continue harmful training or deployment.
};

\draw[arrow] (ai) -- (externalities);
\draw[arrow] (externalities) -- (tax);

\draw[arrow] (tax.south) -- ++(0,-0.25) -| (purposes.north);
\draw[arrow] (tax.south) -- ++(0,-0.25) -| (institutional.north);

\coordinate (merge) at ($(purposes.south)!0.5!(institutional.south) + (0,-0.35)$);

\draw[thick] (purposes.south) -- ++(0,-0.35) -- (merge);
\draw[thick] (institutional.south) -- ++(0,-0.35) -- (merge);
\draw[arrow] (merge) -- (welfare.north);

\draw[arrow] (welfare.south) -- ++(0,-0.25) -| (design.north);
\draw[arrow] (welfare.south) -- ++(0,-0.25) -| (pitfalls.north);

\end{tikzpicture}
\end{adjustbox}

\caption{Summary of the paper's main findings and arguments}
\label{fig:ai_tax_welfare_channels}
\end{figure}

%% file: sections/case_studies.tex
\section{Introductory Examples}\label{sec:case_studies}

To illustrate the potential application of taxation to AI, we consider two contexts: (1) the effects of data centers on water and energy access for nearby residents, and (2) the displacement of creative labor by generative AI. 
In both cases, 
\emph{AI development generates benefits for firms, while shifting costs onto other agents, suggesting that taxation may be appropriate.}
(Note that this section is brief, and more extended analyses can be found in \Cref{sec:ai_externalities} and \Cref{sec:designing_tax}.)

Our discussion highlights several themes that we return to throughout the paper.
First, taxation is a multifaceted regulatory instrument, whether the goal is to discourage unwanted behavior, redistribute technological benefits, or generate revenue needed for other regulatory proposals.
Second, taxation is a blunt instrument that is highly effective at shaping behaviors. Yet for the very same reason, taxation requires careful design to avoid unintended consequences, including burdening actors it did not intend to target or failing to close loopholes that drive even more harmful activity. 
Third, there are several hurdles to taxation, such as the measurability and traceability of the taxable activity as well as defining the tax base and choosing the instrument.
We include more in-depth discussions of instruments in \Cref{ssec:main_tax_instruments,sec:designing_tax}, and of 
externalities in \Cref{sec:ai_externalities},
and design choices in \Cref{ssec:design_choices}.

\subsection{Example 1: Rising water and energy costs for residents}\label{ssec:example_energy_water}

Computing (or, more colloquially, ``compute'') is one of the main ingredients of AI.
Hardware like graphics processing units (GPUs) are necessary not only for training AI models on large amounts of data, but also for running inference each time a user submits a query. Because compute is the backbone of AI{~\cite{sevilla2022compute}}, companies (and even nation-states) are engaged in a race to build and house more of it in what are often referred to as ``data centers''. This race is reflected in rising global demand for data center capacity and large-scale AI infrastructure investments in the U.S., China, and elsewhere~\cite{jin2020energy, iea2025energyai, chen2025ai}.

An increasingly important fact is that data centers require energy to run and water for cooling. The fast expansion of hyperscale data centers has brought with it \emph{higher electricity and water charges for residents living near them}~\cite{kay2026processing,han2026small}. In Ohio's data center corridor, AEP Ohio announced that residential customers should expect an electricity bill increase of \$27 in June 2025, partly due to rising demand from data centers~\cite{aepohio2025electric_bill_adjustments, whoriskey2025aiexplosion}. Moreover, accounts from communities hosting data centers suggest that localized changes in water pressure can create supply interruptions~\citep{tan2025metawatertaps}.
Further, the increased demand strains local supplies in ways that often \emph{require upgrades to infrastructure that are financed locally}, often by taxpayers~\cite{han2026small, martin2025extracting}.

The rising costs are due not only to rising demand, but also to demand uncertainty~\cite{martin2025extracting, satchwell2025largeloads}. For instance, utilities must anticipate demand from both residents and data centers~\cite{klass2026allocating, pjm2025_load_adjustment}. When the realized usage is lower than the expected demand, utilities may, subject to regulatory approval, recover their costs by \emph{redistributing them  across different customer classes}~\cite{naruc2026ratemaking, wood2016recovery_fixed_costs}. Thus, even if residents usage remains constant, their bills could increase due to the uncertainty and magnitude of data center demand. 

\paragraph{Policy responses and role of taxation.}
Recent policy responses in Ohio and across the U.S. have begun to recognize the issue~\cite{columbus2026datacenterhearing}. 
In Ohio, the Public Utilities Commission's ``data center tariff'' reduces risks of speculative data center grid costs from being shifted onto other ratepayers. The tariff~\cite{occ2026quick_facts_data_centers_ohio, aepohio2026ratesettlement} requires large data centers to pay for at least 85\% of contracted capacity for up to 12 years (even if their consumption is lower), though critics point at the protection's flaws~\cite{oma2026inflated_load_forecasts}.  

Taxation could serve as a powerful revenue-raising and corrective instrument here. While taxation may not reduce the number of data centers, 
{it can encourage AI companies to look for more energy- and water-efficient solutions.}
For instance, it may \emph{encourage AI companies to develop more efficient training and inference procedures}. 
It may also encourage them to have \emph{more predictable energy usage} to avoid the demand uncertainty described above, or to use recycled instead of potable water~\cite{mytton2021data}. 
At the same time, 
{the revenue generated can be used to offset these burdens}, for example by \emph{funding grid and infrastructure upgrades}, \emph{lowering costs for affected residents}, or supporting \emph{water and environmental protection measures}. 
The revenue generated by a water or energy tax could also be used for other public goods and services, such as the creation and maintenance of a database for chatbot-related mental health incidents.

Designing taxation for this purpose requires care.
One could tax electricity usage, water consumption, revenue, reserved capacity, or a combination of these factors. 
This choice of tax base and corresponding instrument depends on the intended purpose and effects.
For instance, an excise tax on electricity usage encourages AI companies to lower electricity use, while one on reserved capacity incentivizes companies to have more predictable or steady demand. 
Further, the taxed base should be measurable. 
In this context, electricity and water usage are easily measurable, as these numbers are collected by utilities providers.
However, one may wish to distinguish between usage that is AI-related and usage that is not, which is much more challenging~\cite{guan2023wattscope}. 
As an example: is the electricity used to generate search engine results, which indirectly generates data used to train AI models~\cite{ganchev2012using}, considered AI-related?
A related challenge arises in determining the taxpayer. Possibilities include AI developers, AI deployers, or data center owners. 
Choosing among actors along the AI supply chain~\cite{hopkins2025ai, cen2025mapping} requires considering who should bear the economic burden, which tax instrument is most feasible to administer, and whether the combined choice of tax base and taxpayer shapes incentives appropriately.

The unintended effects of taxation are also important: taxing electricity usage creates tax burdens that (approximately) scale with the size of an AI company, whereas the effect of taxing reserved capacity on AI companies is less clear.
Taxing electricity or water usage may also prompt AI companies to ``shop'' for jurisdictions with the lowest taxes,
including offshoring computation.

\subsection{Example 2: Impacts on creative work}

Data is often referred to as ``the lifeblood of AI''~\cite{weaver2017artificial}. 
This dependence is especially pronounced for large ``foundation'' or ``base'' models that are developed for general-purpose use~\cite{bommasani2021opportunities}. 
For instance, 
GPT-3 was trained on  approximately 240 billion unique tokens (equivalent to a few million books), while more recent models like Llama 3 were trained on 15 trillion tokens~\cite{brown2020language, meta2024introducing}.

The data behind AI is often sourced from humans
and, in many cases, the humans behind the data did not consent to the use or receive compensation or credit.
Some of these concerns are reflected in litigation, such as \emph{Andersen v. Stability}~\cite{andersen2024stability_ai_order}, where artists allege that their works were copied without authorization to train AI, and \emph{Bartz v. Anthropic}~\cite{bartz_anthropic_order}, where authors challenged Anthropic's use of copyrighted books (although the court held that their use for model training constituted fair use)~\cite{stober2025generative, lee2023talkin}. 
These cases reflect the sentiment that AI companies \emph{capture economic value from creative works while the creators themselves do not receive a share of the gains enabled by their contribution}. 
Creative works can thus quietly move from being available online, to being commercially used for AI training, without ordinary mechanisms through which creators control or monetize reuse.

More broadly, there are fears that AI systems can substitute for the labor of the very creators whose work helped train them. Studies have shown that AI systems can sometimes reproduce or closely replicate elements of their training data~\cite{somepalli2023diffusion,carlini2023extracting}. Creators allege that this weakens their income opportunities~\cite{jiang2026professional, ortiz2023ai_copyright_testimony}. 
As AI-generated works reduce demand for human creators in the short-term~\cite{hui2024short}, the technology may gradually reorganize or displace them~\cite{erickson2024ai}.
In response, 
many artists and creators have begun exploring ways to protect and preserve their work via data unions~\cite{hardjono2019data}, renegotiated contracts~\cite{bender2025generative}, trusts~\cite{chan2023reclaiming}, and licensing~\cite{usco2025generative_ai_training}. 
We further discuss this externality in \Cref{sec:ai_externalities} as well as potentiall tax solutions in \Cref{sec:designing_tax}.

\paragraph{Policy responses and role of taxation.}
Society may not reach a normative consensus on whether training constitutes fair use or whether AI's effect on creative labor is an acceptable outcome of technological progress anytime soon. 
Thus, legislative changes that hinge on these debates may be slow.
Moreover, many free market solutions (such as data tracing, attribution, and compensation schemes) will require new technologies and infrastructure that take time to develop.  
All the while, generative AI's effect on creators will persist. Further, the creative industry itself may shrink due to creators' fears of future displacement~\cite{jiang2026professional, ortiz2023ai_copyright_testimony}.

What taxation offers is a bridge---a revenue raising tool that does not bar companies from using data or place requirements on them, but rather nudges them to change their data practices.
Precisely what practices are nudged depends on the tax design, e.g., tax exemptions for companies that participate in direct data compensation schemes or that publish their data sources.
Meanwhile, taxation generates revenue that can be used to mitigate the negative externalities of generative AI on creators or to fund regulatory capacity around data regulations. 
For instance, \emph{revenue could support education or upskilling programs} for displaced workers, 
or it could \emph{fund the creation and monitoring of data use registries} (a costly effort).

Designing taxation for this purpose requires care.
Compared to energy and water use, 
creative labor externalities are harder to trace and measure:
the problem is not simply the volume of AI activity, but the quantification of ``how much'' creative data an AI system uses and 
the effect across many dispersed creators.
Rather than tracing generative AI's effect on creators,
one could tax more easily measurable activities. 
It could be based on training procedures (e.g., the number of training data samples that are not synthetic) or on profits due to model development. 
Policymakers could further offer credits or exemptions based on whether a firm compensates creators or whether they disclose their training data sources.
Note that proxy measures matter: credits based on how much a firm compensates creators versus how many creators it compensates both appear to target the same harm, but they incentivize very different behaviors (large licensing deals versus many small payments).
These design choices depend on the intended purpose of taxation.
For instance, taxing profits from generative AI products raises funds for redistribution but may do little to change data practices because it does not target data practices, while exemptions tied to compensation schemes more directly reshape incentives.

Defining who should pay the tax is similarly difficult. 
Creative externalities are tied to intangible activities distributed across the AI supply chain~\cite{hopkins2025ai, cen2025mapping}, from those who assemble training datasets to those who release generative AI tools. 
A tax aimed at training data collectors may not force generative AI companies that most profit from human data to internalize the harm they cause, while a tax on deployers only indirectly penalizes the harmful activity (e.g., scraping) and leaves data collectors untouched.

%% file: sections/background.tex
\section{Background on Taxation}\label{sec:background}

We first provide the conceptual foundation for the paper's later analysis of AI taxation. Before assessing whether taxes can address AI externalities, we give a high-level overview on what taxes are, why governments use them, and how they operate as economic and regulatory instruments.

\subsection{Basics and key terminology}\label{ssec:background_basics}

Taxes long predate modern welfare states~\cite{jursa_moreno_garcia2015ancient_near_east_egypt}, having served as a means for political authorities to raise revenue, sustain collective power, and administer society. 
Still today, the underlying mechanism has not changed much: taxes continue to transfer resources from private actors to political authorities. In line with this, the OECD's modern definition for taxation consists of ``compulsory, unrequited payments to the general government or to a supranational authority''~\cite{oecd_revenue_statistics_2025_executive_summary}.

\input{sections/table_definitions_terms}

Importantly, taxation carries multiple purposes, which are not mutually exclusive but are vital to the tax design. The tax system generally has three purposes in modern literature: \textbf{(1) raising revenue, (2) redistributing income, and (3) shaping taxpayer behavior}~\cite{harpaz2026taxingai}. The first, raising revenue, finances government services.
For instance, U.S. fuel taxes go toward the Highway Trust Fund, which is the main source of funding for highway infrastructure in the U.S.~\cite{dumortier2017state}. The second goal behind the tax system is redistribution, meaning the use of taxes to rebalance wealth~\cite{OECDcausa2017redistribution}. Typically, redistribution occurs either through direct cash transfers or through in-kind provision of goods and services, especially in education or health services~\cite{currie_gahvari2007transfers_cash_kind}. 
Finally, taxes can shape behavior through both ``positive and negative incentives'', with tax benefits that encourage favored activities and tax burdens that discourage socially undesirable ones~\cite{harpaz2026taxingai}. These purposes are not mutually exclusive, but the intended objective of taxation is critical to its design and implementation, as it shapes the chosen scope, force, and instrument.

Translating these purposes into policy requires clarity about how a tax is structured. Several concepts recur throughout this paper.
The first distinction is between \emph{what} is taxed and \emph{who} pays. The \textbf{tax base} is the economic object to which a tax applies~\cite{stewart_2022, slemrod2014insights}---for example, income, profits, consumption, property, electricity use, or a particular transaction. 
The \textbf{taxpayer}, by contrast, is the party on whom the law imposes tax liability, while the \textbf{tax remitter}, is the party legally required to transfer payment to the government~\cite{slemrod2008does}. These can be the same entity, but need not be. For instance, sales tax is placed on the purchaser, but it remitted by the seller.
Similarly, a tax may target electricity consumption while assigning remittance to a data center owner, an AI developer, or a deployer.

A second set of choices concerns \emph{how much} is owed and \emph{in what legal form} the tax is imposed. The \textbf{tax rate} specifies the charge per unit of the base, whether as a percentage of value (an {ad valorem} rate) or a fixed amount per unit (a {per-unit} charge)~\cite{keen1998balance}. 
The \textbf{tax instrument} refers the legal form of the levy---such as an income, excise, or payroll tax. Two taxes may therefore share a base but differ in instrument, or use the same instrument with different bases. 
Finally, \textbf{tax incidence} describes who ultimately bears the economic burden of a tax after prices, wages, and contracts adjust, regardless of who is legally liable or remits payment~\cite{fullerton2002tax, seligman1927shifting_incidence_taxation}. 
The legal taxpayer and the actor who absorbs the cost need not be the same. 
For example, taxpayers may pass costs 
onto consumer via higher prices; 
however, as we note in \Cref
{ssec:background_counterarguments}, passing costs to the 
consumer can be consistent with a corrective tax because it can therefore reduce demand for the harmful activity by pricing that harm.

\subsection{Economic perspectives on taxation and externalities}\label{ssec:economic_bg}

Taxation is, of course, highly debated. Positions on taxes are central to elections and to broader fights over the size and role of government. That political contention reflects a longstanding debate over taxation's purpose and effects. 
Classical economic theory, for instance, treats taxation primarily as a \emph{market distortion}. 
That is, taxes disrupt what would otherwise be an \emph{efficient market}.
When markets are efficient, market forces keep harmful activities in check because prices would reflect the social costs associated with harmful activities and thus drive demand for those activities down.
Under this view, taxation is a necessary evil to fund the government and meet other fiscal obligations, so the goal is to identify the minimal taxation needed. 

An alternative view was offered by Arthur Pigou in 1920 \cite{pigou1920economics_welfare}---that, in practice, ``market failures'' arise because not all costs associated with an activity are internalized by the actors.
Specifically, not social costs are reflected in prices, 
and the market thus cannot account for them,
which motivates the need for \emph{regulation}. 
Taxation, then, can be \emph{corrective} in that it serves a regulatory purpose that ``corrects'' the market failure by forcing actors to price in the social costs of their activity.
\Cref{fig:ai_tax_starting_point} summarizes these dynamics.

More broadly, an \emph{externality} arises when there are  ``spillover'' effects from an activity~\cite{lemieux2021threat_externalities} that result in uncompensated costs imposed on third parties, where these costs are not fully internalized by the agent who generate them or reflected by market prices~\cite{paniagua2024externalities}.
Note that we use ``externalities'' to refer to negative externalities, 
though it could also include positive externalities (beneficial spillover effects). 
While Pigou did not use the term ``externality'' to describe this divergence between agents' private and social costs~\cite{boudreaux2019externality}, his analysis captures the concept as it is now understood. 
Pigou illustrated negative externalities through examples such as industrial smoke pollution, which leads to environmental degradation, and positive externalities through knowledge spillover, which enhance social welfare~\cite{pigou1920economics_welfare}. He justified the use of taxes, in appropriate circumstances, as a way to correct market failures.

While ``Pigouvian'' corrective taxation is widely used by governments as a way to correct technological externalities, it has its own implementation and scope limitations. For instance, the literature warns that corrective taxation is ineffective when social costs are difficult to estimate~\cite{fleischer2015curb} or when the state cannot condition on all the variables that affect expected harm~\cite{shavell2010corrective_taxation_liability}. Although Pigouvian taxation is justification for taxation, this paper explores different objectives of and approaches to taxation. 
For instance, non-Pigouvian tax instruments can indirectly address technological externalities by generating public revenue instead of compelling actors to internalize the social costs of their activities. 
Rather than pricing the externality itself, the public revenue from taxation can strengthen the regulatory capacity needed to implement and enforce other policies,
or it can support redistribution (e.g., cash transfers or in-kind social services~\cite{currie_gahvari2007transfers_cash_kind}).

\input{sections/figure_tax_plot.tex}
\FloatBarrier

\subsection{Main tax instruments}\label{ssec:main_tax_instruments}

In this section, we survey several existing instruments.
We discuss possible AI tax instruments in \Cref{sec:designing_tax}.
\begin{enumerate}[itemsep=0pt]
    \item \textbf{Corporate income taxes.}
    Corporate income taxes are placed on the net profits of businesses, where profits are given by revenue minus expenses. 
    Their primary purpose is usually revenue-raising rather than externality-correcting. 
    Corporate income taxes allow governments to capture some of the economic gains earned by firms. 
    The resulting tax revenue may then be used to fund regulatory capacity or redistributive efforts. %
    Because corporate income taxes apply to profits broadly, they do not directly distinguish between socially beneficial and harmful gains.

    \item \textbf{Excess-profit, windfall-profit, and rent taxes.}
    These taxes are placed on returns that exceed some threshold of ordinary profit. 
    These taxes are designed to capture unusually high gains, sometimes referred to as ``super profits'' or ``economic rents''~\cite{garnaut2010principles}. 
    Such gains may arise from unusual market power, scarcity, or crises that allow a firm's profits to rise far above a normal competitive level. 
    These taxes are usually not corrective in the Pigouvian sense.
    However, they can be used to target concentrated gains, especially if those gains are viewed as dependent on public infrastructure, publicly funded research, or broader social conditions.

    \item \textbf{Consumption taxes.}
    Consumption taxes are applied to the purchase of goods and services. 
    Common examples include sales taxes and Value Added Taxes (VAT)~\cite{bunn2021consumption}. 
    These taxes are typically collected by sellers and remitted to the government, though some of their cost may be passed on to consumers through higher prices. 
    Consumption taxes are primarily intended to raise revenue by taxing spending instead of income, wealth, or profits. 
    Via prices, they may affect demand for taxed goods and services.

    \item \textbf{Excise and Pigouvian taxes.}
    Excise taxes are narrower taxes applied to specific goods, services, inputs, or activities. 
    While consumption taxes are broad, excise taxes are targeted. 
    Historically, they have applied to tobacco, carbon emissions, and gambling~\cite{boesen2021excise}. They are often charged per unit, as a percentage of the transaction/value or via a fee. 
    Because they can be directed at a specific activity, excise taxes are often used as corrective Pigouvian instruments aimed at a social harm. 
    Although they do raise revenue, they increase the price of a harmful activity, 
    thus incentivizing a reduction in the activity or a substitution with a less harmful alternative.

    \item \textbf{Payroll taxes.}
    Payroll taxes are placed on wages and salaries, distinct from income taxes.
    Both are generally collected through withholding, 
    but payroll taxes are fixed and include contributions from both employers and employees while income taxes follow a progressive rate structure paid by individuals. 
    Moreover, payroll taxes typically fund social welfare programs (such as retirement, healthcare, or disability benefits)~\cite{cilluffo2022payrolltaxes}, 
    whereas income taxes go toward government revenue. 
    Note that when firms replace workers with automated technologies, 
    they may reduce their payroll tax obligations, as these taxes currently only apply to human labor~\cite{mazur2018taxing}. 

    \item \textbf{Capital, wealth, and asset taxes.}
    These taxes apply to the ownership, transfer, sale, or appreciation of valuable assets or resources~\cite{drometer2018wealth}. 
    They include property, estate, capital gains, and net wealth taxes. 
    These instruments are typically used for revenue raising and redistribution~\cite{davies1991distributive}. They are especially relevant where technological change increases the value of capital assets, such as shares in technology companies, intellectual property, land, or infrastructure. Because gains from technologies may accrue disproportionately (e.g., to founders, shareholders, investors), 
    taxes on capital and wealth can be used to redistribute those gains more broadly.

    \item \textbf{Environmental or ``green'' taxes.}
    Environmental taxes apply to activities that create harms to the environment. 
    For instance, carbon taxes are paid by companies based on the amount of greenhouse gases they emit,
    which is intended to force companies to internalize the social cost of their effect on the environment. 
    Green taxes are corrective in that they target activities for which the free market price does not reflect the environmental externalities. 
    They are intended to incentivize companies to reduce emissions, though they may pass the cost onto consumers through higher prices. 
\end{enumerate}
This list is not exhaustive and other forms of taxation exist, such as tariffs and personal income taxes.

\subsection{Challenges}\label{ssec:challenges}

There are notable challenges to designing and carrying out taxation that we briefly survey below.

The first set of challenges arise around \textbf{tax design}---namely determining the tax base, taxpayer, tax rate, and tax instrument~\cite{slemrod2013tax}. 
For instance, policymakers must carefully identify and define the tax base (economic object being taxed) and the taxpayer (actor who pays the tax).
Failing to do so may unintentionally include parties that the government does not intend to tax, or vice versa, and distort incentives.

There is a related concern around tax \textbf{incidence} and how the burden is ultimately allocated. In fact, the actor who pays the tax is not always the party that ultimately bears its economic cost. Depending on elasticity and market structure~\cite{seligman1927shifting_incidence_taxation, fullerton2002tax}, firms may pass down the burden onto buyers through, for example, higher prices (though this effect may be intended, as it can reduce demand for the harmful activity).

\textbf{Legal feasibility and enforcement}~\cite{slemrod2019tax} raise further concerns. Taxed actors may exploit loopholes, restructure their activities, or move operations outside the jurisdiction imposing the tax~\cite{bonaparte2026tax_robots, devereux1998taxes}. These risks are especially significant where the taxed activity is mobile, borderless, or intangible (which holds especially true for technologies). Further, enforcement depends on infrastructure that requires expertise or may be costly, including the tax audit infrastructure.

An important class of of feasibility and enforcement challenges arise around \textbf{measurement, traceability, and valuation}.
For instance, a corrective tax is applied on a measurable activity or, alternatively, measurable proxies for the activity that generates the harm~\cite{shavell2010corrective_taxation_liability}.
Ineffective proxies may fail to track the intended externality, which can lead to miscalibration: undertaxation, where harmful activity remains insufficiently discouraged, or overtaxation, where socially beneficial activity is unnecessarily affected.

\subsection{Counterarguments}\label{ssec:background_counterarguments}

Beyond these design challenges, critics raise broader counterarguments.
In addition to the debate over whether taxation that is discussed in \Cref{ssec:economic_bg},
we enumerate broader counterarguments below:

\begin{enumerate}[itemsep=0pt]
    \item One argument is that taxation does not force actors to stop engaging in the harmful activity. Unlike prohibitions or mandatory requirements, a tax prices harmful activity rather than forbidding it. Firms may treat the tax as a fee and continue the underlying conduct so long as payment remains profitable~\cite{gneezy2000fine}.
    However, this attribute is both a strength and a weakness, 
    as it prices an externality rather than directly forbidding it, which may be too strong of a regulatory action to take. 
    Further, this is only an issue if one views taxation as corrective
    because its revenue-raising and redistributive abilities hold regardless of this counterargument. 

    \item Corrective taxes may require inaccessible information. A Pigouvian tax is ideally set equal to the marginal external cost of the externality, but that cost is often difficult to estimate in practice~\cite{fleischer2015curb} or to condition on all relevant variables that affect expected harm~\cite{shavell2010corrective_taxation_liability}.
    This concern is important if a tax is intended to be corrective and price an externality precisely, but governments can otherwise underestimate the exact price to mitigate market distortion and still collect tax revenue.

    \item The revenue that taxation raises does not automatically benefit those harmed or fund mitigation. Unless the statute explicitly earmarks funds, revenue can be allocated to unrelated public spending rather than compensation, oversight, or infrastructure for affected communities.
    This can be both a benefit and a limitation. Earmarking it ensures that it is used to address a specific harm, but it also limits flexibility of funds, which may be used to address other changing needs.

    \item As discussed in \Cref{ssec:challenges}, the tax burden may fall on the wrong parties rather than the intended ones, which is known as ``incidence''.
    For instance, consumers may ultimately bear much of the cost~\cite{abbott_bogenschneider2018robots_taxes}.
    However, higher consumer prices are often the mechanism through which a corrective tax works, since higher prices can reduce demand. 
\end{enumerate}
We discuss AI-specific challenges, considerations, and counterarguments in \Cref{sec:pitfalls}.

%% file: sections/table_definitions_terms.tex
\begin{table}[tbp]
\centering
\footnotesize
\renewcommand{\arraystretch}{1.2}
\newcommand{\taxtabcolwA}{0.1\textwidth} %
\newcommand{\taxtabcolwB}{0.22\textwidth} %
\newcommand{\taxtabcolwC}{0.28\textwidth} %
\newcommand{\taxtabcolwD}{0.32\textwidth} %
\caption{Core tax design concepts and their relation to AI}
\label{tab:tax_design_concepts}
\begin{tabular}{@{}%
  >{\raggedright\arraybackslash}p{\taxtabcolwA}%
  >{\raggedright\arraybackslash}p{\taxtabcolwB}%
  >{\raggedright\arraybackslash}p{\taxtabcolwC}%
  >{\raggedright\arraybackslash}p{\taxtabcolwD}@{}}
\toprule
\textbf{Concept} & \textbf{Meaning} & \textbf{AI-tax example} & \textbf{Why it matters} \\
\midrule

Policy objective 
& The goal the tax is designed to serve. 
& Funding AI regulatory capacity, redistributing gains and costs of AI, pricing cost of data centers' electricity and water use. 
& Shapes all subsequent design choices. \\

Revenue use 
& How collected tax revenue is allocated. 
& Worker transition programs, grid and water upgrades, education, AI regulatory capacity, or general government revenue. 
& A major advantage of taxation is its ability to fund regulatory capacity and redistribute costs/gains. If desired, tax revenue should be earmarked.\\

Tax base 
& The activity, asset, transaction, etc. on which the tax is imposed. 
& Electricity use, water use, tokens, compute, AI-related profits. 
& Determines what is taxed and what is not. Can closely track the externalities or measure an AI activity broadly. Should be measurable and verifiable. Strongly shapes incentives.\\

Taxpayer 
& The actor legally liable for the tax. 
& AI developer, deployer, data center operator, cloud provider, or downstream user. 
& Determines who is formally targeted by the law. \\

Remitter 
& The actor responsible for collecting and transferring the tax to the government. 
& A data center operator may collect taxes from users to remit to government, or AI model provider collect tax from API users to remit to the government.
& Affects administrative process, feasibility, and enforcement. \\

Tax rate 
& The amount charged per unit of the tax base. 
& Dollars per kWh, dollars per gallon of water, percentage of API cost, percentage of AI-related profit. 
& For corrective taxes, should approximate the marginal external cost of the harmful activity, and shapes strength of the behavioral incentive. Otherwise, determines the amount of revenue raised.\\

Tax instrument 
& The legal form through which the tax is imposed. 
& Excise tax, consumption tax, corporate income tax, rent tax, payroll tax, or regulatory user fee. 
& Main structural decision that shapes how a tax is imposed and administered. Determines what existing infrastructure may apply. \\

Incidence 
& The actor who ultimately bears the economic burden of the tax, regardless of who formally pays it. 
& A tax legally imposed on AI firms may be passed on to consumers, enterprise customers, smaller firms.
& Policymakers should check whether possible incidence is acceptable for their policy objective. \\

Exemptions and credits 
& Carveouts or incentives that reduce liability for certain actors or activities. 
& Exempting firms below a certain model-size threshold, crediting firms that meet safety or fairness benchmarks. 
& Allows finer targeting. Adds definition and enforcement complexity, could cause gaming. \\

\bottomrule
\end{tabular}
\end{table}

%% file: sections/figure_tax_plot.tex
\usetikzlibrary{decorations.pathreplacing}
\begin{figure}[tbp]
    \centering
    \resizebox{\textwidth}{!}{%
    \begin{tikzpicture}[font=\footnotesize]

    \begin{scope}[shift={(0,0)}, x=1.12cm, y=1.15cm, >=Stealth]
        \def\axismax{5}
        \def\xmin{0.55}
        \def\xmax{4.45}
        \def\ymin{0.45}
        \def\ymax{4.35}
        \def\shift{1.1}
        \def\arrowgap{0.14}
        \def\arrowyoff{0.04}

        \pgfmathsetmacro{\demandintercept}{\ymax + \xmin}
        \pgfmathsetmacro{\supplyintercept}{\ymin - \xmin}
        \pgfmathsetmacro{\eqx}{(\demandintercept - \supplyintercept)/2}
        \pgfmathsetmacro{\eqy}{\demandintercept - \eqx}
        \pgfmathsetmacro{\eqxSupply}{(\demandintercept - \supplyintercept - \shift)/2}
        \pgfmathsetmacro{\eqySupply}{\demandintercept - \eqxSupply}

        \node[anchor=south, align=center, text width=5.75cm] at (2.5,5.62) {\scriptsize
          A tax shifts supply upward, raising price and lowering quantity of the activity.};

        \draw[->, thick] (0,0) -- (\axismax,0);
        \node[anchor=north west, align=left] at (\axismax-0.75,-0.1) {Quantity of\\AI activity};
        \draw[->, thick] (0,0) -- (0,\axismax) node[above] {Price};

        \draw[very thick, cyan!70!black]
          (\xmin,{\ymin + \shift}) -- ({\ymax - \shift},\ymax);
        \node[cyan!70!black, anchor=south west, align=left] at ({\ymax - \shift - 0.85},\ymax+0.05) {Supply\\ (with tax)};

        \draw[very thick, black] (\xmin,\ymax) -- (\xmax,\ymin);
        \node[black, anchor=south east] at (\xmin+0.75,\ymax+0.1) {Demand};

        \draw[very thick, orange!85!black] (\xmin,\ymin) -- (\xmax,\ymax);
        \node[orange!85!black, anchor=south west, align=left] at (\xmax-0.55,\ymax+0.05) {Supply\\ (without tax)};

        \foreach \ax in {2.72, 3.32} {
          \pgfmathsetmacro{\ylow}{\ax + \supplyintercept + \arrowgap + \arrowyoff}
          \pgfmathsetmacro{\yhigh}{\ax + \supplyintercept + \shift - \arrowgap + \arrowyoff}
          \draw[->, very thick, cyan!70!black] (\ax,\ylow) -- (\ax,\yhigh);
        }

        \node[cyan!70!black, anchor=west, align=left, text width=2.4cm] at (3.25,2.65) {\scriptsize
          \textbf{With a tax}\\(higher price,\\lower quantity)};

        \fill (\eqx,\eqy) circle (2pt);
        \draw[densely dashed] (0,\eqy) -- (\eqx,\eqy) -- (\eqx,0);

        \fill[cyan!70!black] (\eqxSupply,\eqySupply) circle (2pt);
        \draw[densely dashed, cyan!70!black]
          (0,\eqySupply) -- (\eqxSupply,\eqySupply) -- (\eqxSupply,0);

        \draw[->, thick] (-0.35,\eqy) -- (-0.35,\eqySupply);
        \node[left, align=right] at (-0.38,{(\eqy+\eqySupply)/2}) {price\\increase};

        \draw[->, thick] (\eqx,-0.35) -- (\eqxSupply,-0.35);
        \node[below] at ({(\eqx+\eqxSupply)/2},-0.38) {quantity decrease};
    \end{scope}

    \begin{scope}[shift={(8.2,0)}, x=1.12cm, y=1.15cm, >=Stealth]
        \def\axismax{5}
        \def\xmin{0.55}
        \def\xmax{4.45}
        \def\ymin{0.45}
        \def\ymax{4.35}
        \def\shift{1.1}
        \def\arrowgap{0.14}
        \def\arrowyoff{0.04}

        \pgfmathsetmacro{\demandintercept}{\ymax + \xmin}
        \pgfmathsetmacro{\supplyintercept}{\ymin - \xmin}
        \pgfmathsetmacro{\eqx}{(\demandintercept - \supplyintercept)/2}
        \pgfmathsetmacro{\eqy}{\demandintercept - \eqx}
        \pgfmathsetmacro{\eqxSupply}{(\demandintercept - \supplyintercept - \shift)/2}
        \pgfmathsetmacro{\eqySupply}{\demandintercept - \eqxSupply}
        \pgfmathsetmacro{\producerprice}{\eqxSupply + \supplyintercept}
        \pgfmathsetmacro{\taxrevx}{\eqxSupply/2}
        \pgfmathsetmacro{\taxrevy}{(\producerprice+\eqySupply)/2}

        \node[anchor=south, align=center, text width=5.75cm] at (2.5,5.62) {\scriptsize
          A tax generates revenue for redistribution, regulatory capacity, or government function.};

        \draw[->, thick] (0,0) -- (\axismax,0);
        \node[anchor=north west, align=left] at (\axismax-0.75,-0.1) {Quantity of\\AI activity};
        \draw[->, thick] (0,0) -- (0,\axismax) node[above] {Price};

        \fill[green!35, opacity=0.45]
          (0,\producerprice) -- (\eqxSupply,\producerprice) --
          (\eqxSupply,\eqySupply) -- (0,\eqySupply) -- cycle;

        \draw[->, thick, green!50!black]
          (-0.2,\taxrevy) -- ({\taxrevx-0.65},\taxrevy);
        \node[left, align=right, green!50!black] at (-0.22,\taxrevy) {\scriptsize tax revenue};

        \draw[very thick, cyan!70!black]
          (\xmin,{\ymin + \shift}) -- ({\ymax - \shift},\ymax);

        \draw[very thick, black] (\xmin,\ymax) -- (\xmax,\ymin);

        \draw[very thick, orange!85!black] (\xmin,\ymin) -- (\xmax,\ymax);

        \draw[densely dotted, thick]
          (0,\producerprice) -- (\eqxSupply,\producerprice) --
          (\eqxSupply,\eqySupply) -- (0,\eqySupply) -- cycle;

    \end{scope}

    \begin{scope}[shift={(16.4,0)}, x=1.12cm, y=1.15cm, >=Stealth]
        \def\axismax{5}
        \def\xmin{0.55}
        \def\xmax{4.45}
        \def\ymin{0.45}
        \def\ymax{4.35}
        \def\shift{1.1}
        \def\arrowgap{0.14}
        \def\arrowyoff{0.04}

        \pgfmathsetmacro{\demandintercept}{\ymax + \xmin}
        \pgfmathsetmacro{\supplyintercept}{\ymin - \xmin}
        \pgfmathsetmacro{\eqx}{(\demandintercept - \supplyintercept)/2}
        \pgfmathsetmacro{\eqy}{\demandintercept - \eqx}
        \pgfmathsetmacro{\eqxSupply}{(\demandintercept - \supplyintercept - \shift)/2}
        \pgfmathsetmacro{\eqySupply}{\demandintercept - \eqxSupply}
        \pgfmathsetmacro{\eqxSoc}{(\demandintercept - \supplyintercept - 2*\shift)/2}
        \pgfmathsetmacro{\eqySoc}{\demandintercept - \eqxSoc}

        \node[anchor=south, align=center, text width=5.75cm] at (2.5,5.62) {\scriptsize
          Pigouvian taxes aim to align supply with marginal social cost.};

        \draw[->, thick] (0,0) -- (\axismax,0);
        \node[anchor=north west, align=left] at (\axismax-0.75,-0.1) {Quantity of\\AI activity};
        \draw[->, thick] (0,0) -- (0,\axismax) node[above] {Price};

        \draw[very thick, red!75!black]
          (\xmin,{\ymin + 2*\shift}) -- ({\ymax - 2*\shift},\ymax);
        \node[red!75!black, anchor=south west, align=left] at ({\ymax - 2*\shift - 0.55},\ymax+0.05)
          {\textbf{True marginal}\\\textbf{social cost}};

        \draw[very thick, cyan!70!black]
          (\xmin,{\ymin + \shift}) -- ({\ymax - \shift},\ymax);

        \def\bracex{2.05}
        \def\braceoff{0.16}
        \pgfmathsetmacro{\yPurpleBrace}{\bracex + \supplyintercept + \shift}
        \pgfmathsetmacro{\yRedBrace}{\bracex + \supplyintercept + 2*\shift}
        \draw[decorate, decoration={brace, amplitude=5pt, mirror}, thick, red!75!black]
          ({\bracex+\braceoff},\yPurpleBrace) -- ({\bracex+\braceoff},\yRedBrace);
        \node[
          red!75!black, anchor=west, align=left, text width=2.15cm,
          fill=white, fill opacity=0.72, inner sep=2.5pt, rounded corners=1.5pt
        ] at ({\bracex+\braceoff+0.36},{(\yPurpleBrace+\yRedBrace)/2})
          {\scriptsize Unpriced gap if tax underestimates marginal external cost};

        \draw[very thick, black] (\xmin,\ymax) -- (\xmax,\ymin);

        \fill[red!75!black] (\eqxSoc,\eqySoc) circle (2pt);
        \draw[densely dashed, red!75!black]
          (0,\eqySoc) -- (\eqxSoc,\eqySoc) -- (\eqxSoc,0);

        \draw[->, thick, red!75!black] (-0.35,\eqySoc) -- (-0.05,\eqySoc);
        \node[left, align=right, red!75!black] at (-0.38,\eqySoc) {\scriptsize socially optimal\\price};

        \draw[->, thick, red!75!black] (\eqxSoc,-0.35) -- (\eqxSoc,-0.05);
        \node[below, align=center, red!75!black] at (\eqxSoc,-0.38) {\scriptsize socially optimal\\quantity};

    \end{scope}

    \end{tikzpicture}%
    }

    \caption{Supply and demand for AI activity under taxation. The left panel shows how a supply-side tax shifts equilibrium to a higher price and lower quantity. The center panel highlights the government revenue raised by the tax (green). The right panel illustrates the Pigouvian perspective: marginal social cost lies above private supply, the socially optimal price and quantity occur where demand meets marginal social cost, and a corrective tax should shift supply toward that point. If the tax underestimates marginal external cost, an unpriced externality remains. (The figure depicts a supply-side excise tax. A tax levied downstream, closer to consumers, is often shown as a downward shift in demand with supply unchanged. The quantity of AI activity decreases. Firms sell at a lower price, while consumers pay a higher total price once the tax is included. Thus, a demand-side tax also lowers quantity ``sold'' and reduces the gains for firms.)}
    \label{fig:ai_tax_starting_point}
\end{figure}
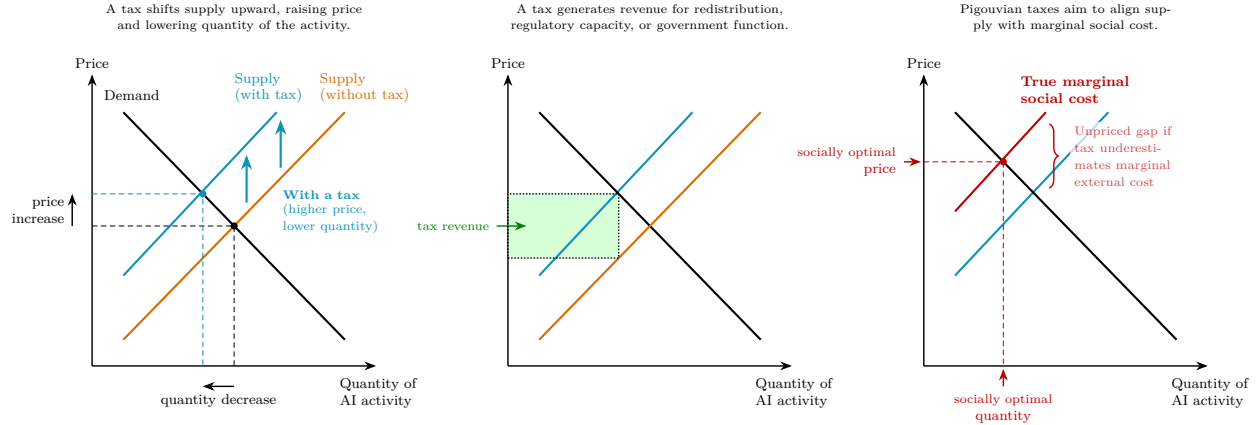

%% file: sections/ai_externalities.tex
\section{AI Externalities}\label{sec:ai_externalities}

\input{sections/table_ai_externalities}

In this section, we discusss potential externalities that an AI tax might target, whether taxation is primarily intended to be corrective or revenue-raising.
For each setting, our goal is to interrogate the question: \textbf{does this activity create negative externalities?}
Although a canonical definition is not universally agreed upon, externalities are generally viewed as \emph{uncompensated costs imposed on third parties that are not reflected in market prices or fully borne by the actors who generate them}.

\subsection{Rising water and electricity prices}\label{ssec:example_externalities_utilities}

As discussed in \Cref{ssec:example_energy_water},
AI computation is primarily housed in what are known as ``data centers''.
These data centers require energy to run and water to keep servers cool. 
Although studies on data centers' energy and water use have been documented for decades~\cite{epa2007data_center_energy_efficiency,sharma2008water_efficiency_datacenters}, 
the recent AI boom has brought critical attention to their rising resource consumption.
This demand is driven by multiple sources, including the training of large general-purpose models (also known as foundation or base models);
post-training of these models to repurpose them for downstream applications;
and running inference each time a user submits a query.

There is increasing concern that the water and energy consumption of data centers are imposing social costs ultimately borne by the general population, regardless of their AI involvement. 
These effects satisfy two components of the definition of externalities. 
First, \textbf{they are not reflected in market prices because utilities are generally priced for public use} (e.g., the cost of water is kept low to ensure that all residents have access to clean water). 
Second, they are not fully borne by the actors who generate them, including AI developers, deployers, and users. 
Data center effects are \textbf{highly localized while the demand for AI that drives data centers is widespread and distant}.

Specifically, 
since water is required for cooling, data centers generally demand the same \emph{unheated, potable water} that the public uses {~\cite{rogers2002water}}.
This overlap might not only result in higher prices for residents, 
but also water quality issues and water shortages \citep{tan2025metawatertaps, fleury_jimenez2025life_next_to_data_centre}.
Further, unheated water is kept at a low price because access to clean water is widely considered a human right. 
Under this perspective, water should not have a market value, i.e., \emph{water is not a traded good} \cite{wheeler_nauges_grafton2023water_pricing}
and should only be priced high enough to  ``recover the costs of the construction, maintenance, and operation of the water delivery infrastructure''~\cite{megdal2005water_pricing_conservation}.
This principle is compromised by allowing residents to pay for \emph{private sector} water usage via rising prices.

A similar but distinct trend holds for data centers' \emph{electricty} demand.
Data centers' electricity consumption is set to double between 2024 and 2030 \citep{iea2025energyai,iea2025aidriver}, resulting in higher electricity prices (there are reports of prices rising 267\% in the past five years \citep{bloomberg2025ai_data_centers_power_bills}). 
Moreover, as discussed in \Cref{ssec:example_energy_water}, rising costs are also tied to \emph{uncertainty} in demand. Since data center usage varies
widely across time, energy providers must anticipate demand. When actual usage falls below anticipated demand, utilities may still recover the costs of planned capacity. Subject to regulatory approval, \emph{those costs can be spread across customer classes, including residents, even if their own usage remains  unchanged}~\citep{naruc2026ratemaking,wood2016recovery_fixed_costs}. 

In addition to the trends above, two additional factors suggest that data centers generate externalities. 
First, on top of rising utility prices, data center demand strains the energy and water grid,
resulting in \emph{costly infrastructure updates} that are typically paid for by the taxes of residents. 
Second, the \emph{effects are distributed unequally} (more often in rural, often low-income residents), 
even when the demand (from developers training AI, deployers applying AI, and even users of AI) is widespread and distant \citep{andthewest2025ai_data_centers_west}.

Thus, the rising energy and water consumption from data centers translate to rising costs for nearby residents, 
either directly via prices or indirectly through, e.g., upgrades to public infrastructure, that are not borne evenly or by the actors who create them, which is strongly suggestive of negative externalities.

\subsection{Labor displacement}\label{ssec:labor_externality}

AI is increasingly used or designed to replace tasks previously completed by humans, not only because it can reduce costs and increase productivity, but because it extends automation to non-routine and creative cognitive tasks \cite{eloundou2024gpts, webb2019impact}. This has raised concerns around human labor displacement across industries, from creative work to customer service industries to software engineering. Scholars and industry leaders increasingly highlight that highly-skilled entry-level roles may especially be exposed \cite{brynjolfsson_chandar_chen2025canaries, openai_industrial_policy_intelligence_age, amodei2024adolescence_technology, vandehei_allen2025white_collar_bloodbath, eloundou2024gpts, economist2025ai_replace_junior_workers}.
    
On a macroscale, labor substitution is not itself an externality, but its widespread effects may generate broader spillovers. First, \textbf{AI may depreciate the value of the same underlying skills across multiple occupations and industries}, which weakens job switching \cite{amodei2024adolescence_technology}. Suppose that a significant fraction of AI-exposed workers are displaced by AI: the demand for their skills will fall while the supply of those skills remains, causing a labor market mismatch \cite{ooi_goh2022taxation_automation_ai}. This results in \textbf{unstable income sources for AI-substituted workers, whose purchasing power decreases, which affects consumption and increases reliance on social support}. As a result, displacement destablizes labor markets, public finance and welfare, while leaving accumulated human capital underused.

This has become a focal point of debate about the creative industry in particular, where the externality is more visible. Creators are not compensated for the creative contents that serve AI training, while AI systems can produce outputs that imitate and compete with those original works \cite{pasquale_sun2025consent_compensation, usco2025generative_ai_training, brauneis2025copyright_training,carlini2023extracting,somepalli2023diffusion, somepalli2024measuring}. As a result, AI activity---namely the training and use of generative AI---compromises intellectual property. This is separate from the fair use debate, which concerns copyright law violations \cite{usco2025generative_ai_training}. Even where training is ultimately lawful \cite{chhabria2025kadrey_meta_order}, \textbf{AI systems still use intellectual property without attribution or compensation}. This generates externalities for creators: AI may reduce creator income~\cite{jiang2026professional} and weaken incentives for human creation~\cite{brauneis2025copyright_training}. Both affect job security, constituting social costs that are not absorbed by those actors, who profit from the AI activity.

Finally, if \textbf{AI replaces a growing share of workers, it may also reduce public revenue needed to fund government services and social support previously obtained through payroll taxes}~\cite{bearer2025sharing, korinek2026publicfinance}. Unlike human workers, whose earnings are subject to income and payroll taxes~\cite{cbo2026taxes}, AI systems are not taxable persons under existing tax systems \cite{abbott_bogenschneider2018robots_taxes}. Governments may therefore lose part of the tax contributions previously generated by displaced workers. This creates a fiscal spillover. 
Unemployment benefits, retraining, and other forms of social support may become more necessary while labor-based public revenues will decline~\cite{bearer2025sharing, korinek2026publicfinance}. Meanwhile, firms capture gains from AI substitution, such as greater productivity, without bearing the full fiscal and social costs of displacement.

AI displacement matters not only because workers may lose jobs, but because the costs of that transition---the externalities---may be pushed onto workers, public budgets, and the economy more broadly.

\subsection{Other potential externalities}\label{ssec:other_potential_externalities}

The previous subsections provide in-depth discussions of two AI externalities. Below, we briefly list several additional consequences of AI that could be considered externalities, providing a brief discussion of them.\\

\textbf{Environmental costs.} 
Related to our discussion of rising utility costs in \Cref{ssec:example_externalities_utilities}, 
AI infrastructure consumes sizable quantities of environmental inputs, both for (1) AI training and inference and for (2) producing and operating the necessary computing equipment. Market prices fail to reflect the full social costs of carbon emissions \cite{muller2026measuring}, water depletion~\cite{li2025ai_water_footprint}, hardware production and electronic waste~\cite{strubell2019energy, schneider2025life, falk2025flops, xiao2025environmental}. AI firms do not bear the full environmental costs of their operations, shifting the damages onto the public and future generations.
This argument is similar to broader discussions on environmental externalities and ``green'' taxes, 
and one may argue that taxing environmental costs of AI activities is double counting. 
Thus, the goal should therefore be to tax environmental externalities that are not already taxed; for example, by addressing \emph{data centers' externalities on access to affordable and clean water}.

\textbf{Misinformation and hallucinations.} Generative AI can be used to produce persuasive false content~\cite{park2026generative} either intentionally, when malicious users deploy it for disinformation~\cite{romanishyn2025ai}, or unintentionally, when models hallucinate plausible but non-factual claims~\cite{huang2025survey}. Because this content can be produced at low marginal cost~\cite{international_ai_safety_report2025}, AI can increase the supply of misinformation, ultimately polluting shared information environments~\cite{huang2023digitalcommons, zhang2025economics}. As a result, users bear costs both (1) downstream, by correcting decisions based on false AI outputs, and (2) upstream, by spending additional time, money, and institutional resources verifying outputs before use~\cite{catalini2026some, campante2025genai}.
These costs are diffuse and thus difficult to price. 
Moreover, because nearly all generative AI systems carry this risk, they are not accounted for even when there is perfect market competition when the pull of generative AI tooling is high. 

\textbf{Bias and discrimination.} 
AI systems can propagate (and, at times, amplify) biases that exist in their training data~\cite{barocas2016big}. 
Because AI has permeated so many different industries, these social costs arise across hiring~\cite{zhang2024measuring}, lending~\cite{bartlett2022consumer}, healthcare~\cite{obermeyer2019dissecting} and education~\cite{baker2022algorithmic}.
This can cause diffuse, systemic harms that are difficult to trace and price into the market.
Such systemtic harms interfere with efficiency; 
they create consistent and persistent ``errors'' in decision-making and allocation. 
Affected individuals spend resources on resolving frictions to detect, appeal, and correct discriminatory decisions~\cite{hakkarainen2021naming}. 

\textbf{Privacy.} Private information can enter AI systems through (1) training datasets that may contain personal, private, and sensitive information~\cite{carlini2021extracting}, and (2) user interactions logs that also enter the training data. 
In both cases, firms capture free data for model improvement~\cite{king2025user, openai2026data_model_performance}, while individuals may lack consent or control over the inclusion and processing of their data. 
The relative harm of having private, personal, or sensitive information released is much higher for individuals than for firms or their customers, 
and it is in some sense an ``irreversible'' harm.
Therefore, individuals bear the burden of privacy loss~\cite{carlini2021extracting}, with risks of unwanted inferences~\cite{staab2024beyond}, data leakage or identity theft, and financial costs. 
Further, they also bear the burden of requesting that their data be removed or legally contesting the use of their data. 

\textbf{Cybersecurity and system vulnerabilities.} 
Recent frontier models have lowered the effort, cost, and skill threshold for malicious actors to engage in offensive cyber activity \cite{brundage2018malicious, fang2024llm}. 
This creates three social costs: (1) direct losses from attacks, (2) legal burdens, and (3) insurance and future security hardening costs \cite{cobos2024review}. 
When firms release frontier models that make offensive cyber activity easier, the resulting costs fall mainly on victims and institutions rather than the firms or its users, 
and those costs are not fully priced into firms' decisions about model release and access.

\textbf{Catastrophic risk.} 
Catastrophic risks refer to high-impact, irreversible damage due to AI development, such as loss of control (e.g., AI systems turning on humanity) and severe misuse (e.g., use of AI to develop bioweapons that wipe out significant populations)~\cite{international_ai_safety_report2025, hendrycks2023overview,blomquist2025aisafetyglobalpublicgood}.
Catastrophic risks can be considered, in some narrow ways, similar to environmental risks in that there is the possibility of irreversible harm.
Yet, regardless of whether this outcome occurs, the possibility of catastrophic damage imposes costs on society through fear and society's response to that fear. %
Furthermore, AI developers often benefit from claiming their models are more powerful than others, which can feed this fear (though increasing attention is paid to ``safety'').
Firms thus capture the private gains from frontier development, but the costs associated with catastrophic risks  are not priced into their release and scaling decisions~\cite{bostrom2013existential}.
\\

Our (non-exhaustive) analysis of the different externalities that arise from AI development and deployment illustrate a common structural problem. AI firms \emph{capture large private gain from AI activities, while they fail to internalize the full economic and social costs} distributed to different individuals, organizations, and society at large. 
This is, at times, due to the \emph{breadth} with which AI has been adopted as well as the \emph{diffuse} effects of AI that are difficult to trace.  
In fact, AI firms may \emph{benefit} from activities that cause or exacerbate social harms (e.g., releasing powerful models allows them to be applied positively but also maliciously).
Thus, achieving a more efficient market requires pricing in those activities.

%% file: sections/table_ai_externalities.tex
\begin{table}[htbp]
\centering
\footnotesize
\setlength{\tabcolsep}{4pt}
\renewcommand{\arraystretch}{1.25}
\newcommand{\extabcolA}{0.17\textwidth} %
\newcommand{\extabcolB}{0.24\textwidth} %
\newcommand{\extabcolC}{0.28\textwidth} %
\newcommand{\extabcolD}{0.25\textwidth} %
\caption{Potential AI externalities: repercussions, rationale, and measurement challenges}
\label{tab:ai_externalities}
\begin{tabular}{@{}%
  >{\raggedright\arraybackslash}p{\extabcolA}%
  >{\raggedright\arraybackslash}p{\extabcolB}%
  >{\raggedright\arraybackslash}p{\extabcolC}%
  >{\raggedright\arraybackslash}p{\extabcolD}@{}}
\toprule
\textbf{Harm/risk}
& \textbf{Repercussions}
& \textbf{Why a potential externality}
& \textbf{Measurement challenges} \\
\midrule

AI computation demands significant electricity and water.
& Expanding data centers raise local utility prices, strain grids, require infrastructure upgrades often funded by taxpayers, and can create water insecurity.
& Harms are localized while AI demand is widespread and distant; costs fall disproportionately on nearby communities and low-income residents; compromises access to affordable water and electricity.
& Hard to separate AI-related demand from other data center or cloud use; local harms depend on grid capacity, water stress, and infrastructure context. \\
\addlinespace[0.35em]

Firms substitute AI for human labor.
& Job loss and wage pressure, weaker purchasing power, greater reliance on social support, and reduced payroll tax revenue.
& Firms capture productivity gains; displacement and fiscal costs fall on workers and governments rather than AI adopters.
& Hard to distinguish displacement from augmentation or ordinary productivity growth; effects are uneven across occupations and time. \\
\addlinespace[0.35em]

AI training and deployment generate emissions, resource use, and e-waste.
& Higher emissions, water use, pollution, and hardware waste from training, inference, cooling, and chip production.
& Environmental harms are not fully priced into AI activity; costs borne by ecosystems, nearby communities, and future resource users.
& Harms vary by location, energy source, cooling method, and supply chain; difficult to trace AI's share of broader environmental damage. \\
\addlinespace[0.35em]

Generative AI produces misinformation and hallucinations.
& False or unreliable content at scale, downstream verification burden, and eroded trust in shared information environments.
& Content is cheap to produce; verification, correction, and trust-repair costs shift to users, platforms, and public institutions.
& Harms are diffuse and hard to attribute to specific outputs, platforms, or downstream decisions. \\
\addlinespace[0.35em]

AI systems propagate bias and discrimination.
& Wrongful exclusions and errors in hiring, lending, healthcare, and education; appeal and dignity costs for affected individuals.
& Firms benefit from automated decisions; discrimination costs fall on protected groups and are not reflected in market prices.
& Discrimination is systemic and context-dependent; individual harms are hard to observe, quantify, or link to a specific system. \\
\addlinespace[0.35em]

AI systems train and sometimes leak private, personal, and sensitive data.
& Unwanted collection, inference, and reuse of personal data; exposure, breaches, and loss of autonomy for individuals.
& Firms capture data value; asymmetric privacy and security risks borne by individuals who may lack meaningful consent or compensation.
& Privacy loss is hard to value; affected individuals may be hard to identify, and harm is often latent until exposure or misuse. \\
\addlinespace[0.35em]

AI models enable cyberattacks and malicious automation.
& Cheaper cyberattacks, vulnerability discovery, and malicious automation; security losses and hardening costs for victims and institutions.
& Developers and deployers capture gains from powerful models; downstream misuse costs fall on firms, governments, and infrastructure operators.
& Expected harm is uncertain before misuse occurs; hard to forecast which capabilities will be exploited and by whom. \\
\addlinespace[0.35em]

Frontier AI development poses catastrophic risks.
& Exposure to loss-of-control or severe misuse scenarios; precautionary and social costs even when catastrophic harm does not occur.
& Private actors capture frontier gains; low-probability, high-impact risks are broadly distributed across society and future generations.
& Capability, probability, and expected harm are inherently speculative; low-probability events are difficult to quantify ex ante. \\
\addlinespace[0.35em]

\bottomrule
\end{tabular}
\end{table}

%% file: sections/case_for_ai_taxes.tex
\section{The Case for AI Taxes}\label{sec:case_for_AI}

Taxation is a blunt instrument, but its effects are difficult to ignore: by imposing monetary penalties and incentives on economic activity, it reshapes prices and market choices. At the same time, it transfers resources to the public sector that can fund oversight, compensation, or other policy goals.

In this section, we consider why AI taxation may be worth pursuing as a policy response to AI externalities. We organize the argument in two parts. First, we discuss how taxation can support distinct regulatory objectives: shaping conduct, funding oversight, and redistributing gains. Second, we analyze the institutional advantages of taxation, including that taxation has existing infrastructure. 
We discuss the potential pitfalls of taxation as well as open questions in \Cref{sec:pitfalls}.

\input{sections/table_for_against_ai_taxation}

\subsection{Purposes that taxation can serve}

In this subsection, we discuss the objectives that taxation can serve.

\paragraph{Corrective taxation: Taxation can discourage harmful activity by pricing it.}
One purpose taxation can serve is corrective: discouraging harmful activity by pricing it at the margin. 
Pricing harmful activities can trigger three responses (sometimes simultaneously): \emph{(1) the taxpayer reduces the taxed activity to lessen their tax burden; 
(2) the taxpayer substitutes the taxed activity with less harmful alternatives; or 
(3) the taxpayer passes the extra cost to downstream consumers by raising prices.}
For instance, a tax on AI-specific electricity usage might lead a developer to design models to use fewer output tokens when responding as per (1); to replace parts of their AI pipeline with more compute-efficient machine learning algorithms as per (2); or to raise API costs as per (3).

The first and second responses directly advance the corrective goal, as long as substitution in (2) lowers harm on net rather than merely relocates it outside the tax base.
Although the third response raises concerns about incidence (passing costs onto downstream actors rather than the  intended taxpayer), 
\emph{(3) can also be corrective.}
Higher consumer prices (e.g., higher API costs) can lower downstream demand for the harmful activity and therefore indirectly reduce it. 
Further, in response to lower demand, firms may seek out less taxed substitutes, implying that (3) can replicate the effects of (1) and (2).

Note that responses (1), (2), and (3) have different implications. 
Pricing externalities close to downstream demand is more natural when harm scales with downstream use, demand is elastic, and consumers can choose (less harmful) alternatives unaffected by taxation. 
On the other hand, when the externality arises on the production side and harms are less reversible (the fact that the activity occurs is itself harmful), pricing the relevant activity directly and close to the upstream activity is important.
For instance, the harms of electricity demand scale with use, while the harms associated with the unrestricted availability of un-guardrailed frontier models are not reversible.

\paragraph{Regulatory capacity: Taxation can fund the oversight AI governance requires.}
A second purpose taxation can serve is to raise revenue for regulatory capacity. 
AI regulation requires funding, technical expertise, and infrastructure.
For instance, states have passed several laws in recent years, including California's transparency and frontier-model safety legislation~\cite{california2024sb942, california2025sb53} as well as requirements in Colorado, Illinois, and other states~\cite{ncsl2025ailegislation, davtyan2025us}. 
\emph{Effective enforcement depends on the ability to monitor compliance, investigate violations, and adapt oversight as models evolve}~\cite{guha2024ai, jones2020artificial, taxescrscategories}. 

Federal efforts face similar constraints. FY2026 appropriations provided at least \$55 million for NIST's AI research initiatives, with up to \$10 million for the Center for AI Standards and Innovation (CAISI)~\cite{gpo2026hr6938explanatory}; for FY2027, NIST's Congressional budget submission requests \$73 million, including an expanded CAISI effort~\cite{commerce2026nistntisfy2027}. These figures illustrate both the need and the limits of annual appropriations alone.
AI governance requires significant, ongoing funding for monitoring systems, audit capacity~\cite{auditingnyscomptroller2025locallaw144}, risk-management frameworks~\cite{nistrmfprinciplesai2023artificial}, enforcement, and public research infrastructure. 
Taxation offers one way to generate a revenue stream from the sector whose activities create these oversight needs, rather than relying solely on appropriations that rely on political priorities that could shift.

While firms can finance their own compliance programs, credible enforcement also depends on public agencies and shared infrastructure that no single company will fully fund on its own. Tax revenue can therefore support independent, standardized capacity (e.g., standards development, incident-reporting infrastructure, evaluation tools, and auditing capacity) that complements firm-level compliance. Sector-linked financing via taxes also has precedent. FDA tobacco regulation is supported in part through sector-specific user fees imposed on regulated firms~\cite{parasrampuria2023fda, fda_user_fees_explained}. That model differs from a general tax in legal design, but it illustrates how regulated industries can help finance the public oversight their activities require. An AI tax could play a similar role in funding governance infrastructure needed to keep pace with frontier development. Unless a statute earmarks revenue for oversight, funds need not flow directly to mitigation; that design choice is discussed in \Cref{sec:pitfalls}. Even so, \emph{taxation provides a recurring fiscal channel for building the regulatory capacity that existing and proposed AI laws assume but do not themselves fund.}

\paragraph{Redistribution: Taxation can support those who bear AI's costs.}

A third purpose taxation can serve is redistribution. Like regulatory capacity above, this is grounded in taxation's ability to generate revenue, but it redirects it toward distributing the costs of AI rather than at funding oversight mechanisms. 
\emph{Redistribution is motivated by the fact that AI's costs are often borne unequally}. 
For example, data center externalities are primarily experienced by nearby residents via higher utility bills and water insecurity, 
and the creative industry is disproportionately affected by AI's training data practices. 

Although taxation cannot undo these shifts entirely, 
\emph{it can rebalance AI's uneven impacts and, in turn, help society at large prepare for the long-term effects of AI}. 
For example, revenue can be deployed through \emph{targeted} redistribution to directly harmed groups (such as compensation funds for authors of copyrighted works trained on by AI models~\cite{senftleben2023generative} or financial relief for communities facing higher utility bills).
Alternatively, revenue can be deployed through \emph{general-purpose} redistribution that funds broader long-term efforts, including education, retraining programs, and infrastructure upgrades~\cite{unesco2024ai_competency_framework_teachers, pct2018taxation_sdg, makela2024complement}.
These measures may take the form of cash transfers, in-kind services, or social-purpose tax breaks~\cite{currie_gahvari2007transfers_cash_kind, oecd_social_spending_indicator}.
Almost all forms of redistribution require a revenue source, and taxation can provide just that.

Tax-funded social spending has precedent in other sectors. Following California's Proposition 56 in 2016, the state increased tobacco taxes and directed part of the revenue to public programs, including school-based tobacco-use prevention education~\cite{zhu2026california}. It illustrates how revenue from a regulated sector can be directed toward public good and services.
We note that, in order for  AI tax revenue to reach the intended communities or affected individuals, it generally should be earmarked as such. 

\subsection{Institutional and design advantages of taxation}

Beyond the regulatory objectives above, taxation offers institutional advantages.
We highlight several below.
\begin{enumerate}[itemsep=0pt]
    \item \textbf{Easier to pass fiscal legislation.}
    In the U.S.,
    \emph{budget reconciliation} is a process by which Congress can pass major fiscal legislation with a simple majority (instead of the supermajority often required to overcome a filibuster)~\cite{crs2024budget_reconciliation}.
    Tax legislation falls into this exception, 
    allowing it to move through Congress (slightly) more easily than other, non-fiscal regulations. 

    \item \textbf{Existing fiscal, enforcement, and independent audit infrastructure.} 
    There is a longstanding system for assessing, collecting, auditing, and enforcing taxation~\cite{oecd2025tax_administration}. 
    For instance, the tax system includes existing independent public auditors.
    Further, tax authorities have records for many AI firms, against which they can cross-check filings, strengthening enforcement.
    All to say, an AI tax can build on this machinery,
    allowing taxation to be stood up fairly quickly compared to novel oversight programs. 

    \item \textbf{Predictable rules and familiar compliance procedures.} 
    Nearly all firms are familiar with taxation and the procedures surrounding it. 
    Although tax requirements always require some interpretation, firms can turn to a well-established industry of tax accountants~\cite{deluca2007irs_research_conference}. 
    These factors make tax rules more predictable and compliance more straightforward, relative to newer regulatory frameworks.

    \item \textbf{Mitigates disproportionate burden on small firms.}
    In practice, many regulatory approaches to AI governance unintentionally create disproportionately high compliance burdens on small- and medium-sized enterprises because, while larger firms can afford to hire compliance teams, smaller firms must either hire external consultants (costly) or build internal capabilities from scratch (time-consuming).
    By contrast, an AI tax would add new obligations, 
    but it would build on familiar processes that all firms, small and large, already maintain~\cite{deluca2007irs_research_conference}. 

    \item \textbf{Taxation is direct.} 
    Taxation attaches consequences to economic activity.
    By operating in monetary units, it is easy for firms to incorporate taxation into their financial analyses and planning. 
    Firms can therefore reason more concretely about how to change their activities.

    \item \textbf{Design flexibility.} Taxes are a ``flexible policy instrument''~\cite{ciocirlan2003political} that policymakers can adjust through different choices of the tax base, taxpayer, rates, thresholds, exemptions, and credits.
    Taxes can also be engineered to fall primarily on large actors that generate disproportionate harm~\cite{bearer2025sharing}.
    
    \item \textbf{Taxation can shape incentives rather than AI design and definitions.} 
    Taxation works by applying a price to a measurable activity (or a proxy for it). 
    \emph{If} desired, it can price these quantities directly, shaping incentives and behavior without placing specific technical requirements on AI design 
    or defining normative concepts that remain contested. 

\end{enumerate}
We conclude by noting that taxation should be understood as complementary to other AI governance tools, not as a substitute for them. Other proposals may require years to build specialized infrastructure,
develop new standards, and amass domain expertise (for instance, AI auditing is currently undergoing this evolution~\cite{raji2022outsider, mokander2023auditing}). 
While those capacities mature, taxation can help fill governance gaps.

\begin{remark}[Additional governance motivation for taxing AI]
    As examined by others~\cite{harpaz2026taxingai, korinek2026publicfinance, cbo2026taxes},
    the very labor displacement that AI could cause would lead to fewer human employees.
    Because payroll tax revenue only applies to \emph{human} labor, 
    tax revenue will decrease as a result.
    The government would therefore need other sources of revenue.
    Since this displacement is caused by AI, taxing AI-related activities would replace lost tax revenue. 
    This motivation for taxing AI is non-regulatory but serves an important governance need. 
    We discuss this in more detail in \Cref{sec:designing_tax}.
\end{remark}

%% file: sections/table_for_against_ai_taxation.tex
\begin{table}[htbp]
\centering
\footnotesize
\renewcommand{\arraystretch}{1.25}
\caption{Reasons for and against AI taxation (\Cref{sec:case_for_AI,sec:pitfalls}).}
\label{tab:for_against_ai_taxation}
\begin{tabular}{@{}%
  >{\raggedright\arraybackslash}p{0.42\textwidth}%
  >{\raggedright\arraybackslash}p{0.42\textwidth}@{}}
\toprule
\textbf{Reasons for} & \textbf{Reasons against} \\
\midrule

Can price harmful AI activity and shift firms toward less harmful alternatives
& May raise adoption and development costs, slowing innovation and productivity growth \\

Can raise revenue for AI oversight, enforcement, and regulatory capacity
& May weaken competitiveness and encourage relocation to lower-tax jurisdictions \\

Can redistribute gains and support workers and communities bearing AI's costs
& May be weakened by lobbying and political capture during design \\

Builds on existing tax administration, audit, and compliance infrastructure
& AI is hard to define, measure, and link to specific harms or actors \\

Offers flexible design across tax base, rate, taxpayer, and exemptions
& Poor design can create loopholes, threshold distortions, and unintended incidence \\

Can shape incentives without prescribing contested technical definitions of AI
& Firms may treat the tax as a fee to pay rather than a reason to change conduct \\

Can complement other AI governance tools while those capacities mature
& Proxy-based taxes may misprice harms that vary by context or location \\

Can help adapt the tax base as labor-linked public revenue erodes
& May incentivize tax planning and restructuring instead of harm reduction \\

\bottomrule
\end{tabular}
\end{table}

%% file: sections/surveying_options.tex
\section{Designing an AI tax}\label{sec:designing_tax}

There are multiple important decision when designing an AI tax, 
including the policy objective (what the tax should achieve),
the tax base (what to tax), 
the taxpayer and remitter (who owes and pays the tax), 
the tax instrument (how the base is levied), 
the rate and structure (how much is owed), 
exemptions and credits (who is exempt or credited), 
and revenue use (how the revenue is used). 
Because AI activity is distributed across a long supply chain~\cite{hopkins2025ai, cen2025mapping}, where the tax attaches also affects how effectively it reaches the intended harm. 
We discuss these considerations below and illustrate how they might apply to the externalities in \Cref{sec:ai_externalities}.

\subsection{Design choices}\label{ssec:design_choices}

We review the main design decisions below. 

\textbf{Policy objective.} As discussed in \Cref{sec:case_for_AI}, taxation can serve corrective (pricing harmful activity), redistributive (capturing gains for affected parties or the public), and regulatory-capacity (funding oversight) goals.
The policy objective is what the tax design should optimize for. The same externality may motivate more than one objective, but the objective should be clear because it shapes every subsequent choice.

\textbf{Tax base.} The tax base is the activity or quantity being measured and priced. For example, the base could be electricity usage, token volume, or AI-driven profits. 
The base must be observable in that it can be verified based on records, data, or reporting. It must also be aligned with the objective. 

For a corrective tax, the base should approximate the harmful activity, or the closest measurable proxy~\cite{fullerton2002tax}. It should also be a quantity the relevant actor can still reduce, or the tax cannot change behavior at the margin (\Cref{sec:case_for_AI}). For a redistributive tax, the base may instead capture where gains accrue (profits, revenue, or returns above a threshold), even if it does not track each harmful act. For a regulatory-capacity tax, the base may mark the scale or type of activity that creates oversight needs (such as frontier training compute or high-risk deployments). Finally, the scope of the base matters: a narrow base (e.g., electricity at AI data centers) targets the externality more precisely but requires clear boundaries; a broad base (e.g., all cloud compute) is easier to administer but may sweep in unrelated activity.

AI harms are often intangible, so policymakers often tax proxies rather than harm itself, 
but proxy choice requires careful consideration.
Two different bases aimed at the same externality can induce different behaviors, such as when taxing how much a firm pays creators versus how many creators it pays.

\textbf{Taxpayer and remitter.} The \emph{taxpayer} is the party on whom the law imposes tax liability.
The \emph{remitter} is the party legally required to file and transfer payment to the government.
In most of the designs considered here, these are the same entity, but that is not always true (e.g., when a company remits a consumption tax collected from consumers).
In the AI industry,
liability may be assigned to a developer, deployer, data curator, data center owner, or one of the many actors who participates in the AI supply chain~\cite{hopkins2025ai, cen2025mapping, widder2023dislocated, cobbe2023understanding, bommasani2024ecosystem}. 
\emph{The difficulty in AI is that the lines between actors in the AI supply chain are often blurry}. 
For example, most AI products are the result of multiple layers of model training, and almost all downstream companies must post-train upstream models in order to ensure that their products are best suited for their specific context of interest; 
levying a tax on any form of training may encourage downstream companies to use off-the-shelf models and discourage any beneficial post-training.
This example illustrates challenges around appropriately scoping the taxpayer.
Thus carefully defining the taxpayer is a key design choice, though other parameters, such as the tax base and exemptions, can help simultaneously address the same issues.

\textbf{Tax instrument.} 
The instrument is the legal mechanism through which a chosen base is taxed. 
As discussed in \Cref{ssec:main_tax_instruments}, familiar examples include excise, corporate income, and consumption taxes~\cite{due1994excise, crs_r43189}. 

In general, the base and instrument should be chosen together rather than treated as separate, 
and \emph{the choice of instrument should depend on 
(i) the intended purpose, (ii) where an activity most closely attaches to this intended purpose, and (iii) the feasibility of implementing this instrument in practice}. 
For instance, if the purpose is corrective, the base and instrument should track the activity most closely tied to the externality not only to maximize effectiveness, but also to avoid unintended burdens and effects. 
For redistribution or funding regulatory capacity,
rent taxes (levies on profits above a threshold) and
user fees (charges on target firms that are used to fund their oversight) may be more natural.
Finally, instruments that utilize existing administrative infrastructure or build on existing tax code can be more reliable.

Instruments also differ in where they intervene along the AI supply chain~\cite{hopkins2025ai, cen2025mapping}: excises typically target upstream (at production or inputs), whereas consumption taxes attach downstream (at purchase or use of a service). 
These options, which can be deployed together, reflect different policy objectives and externalities. 
For instance, when harm scales with downstream adoption, a consumption tax may be more effective than an upstream excise; the reverse is often true for harms tied to inputs or production, such as the mere fact that a large model is being trained. 
We discuss candidate instruments in \Cref{ssec:ai_instruments}.

\textbf{Rate and structure.} 
The rate matters because it most directly determines how large the tax obligation is. 
There are a few key decision points. 
First, liability could be calculated \emph{per unit} (a fixed amount that is based on quantity, such as dollars per kilowatt-hour), \emph{ad valorem} (a percentage that is based on value and therefore often tied to prices, such as a fraction of sales revenue), or another tax rate schedule. 
Common alternatives include \emph{excess-return or rent-based} structures, in which a tax applies only once activity or returns exceed a predetermined threshold, and \emph{progressive} or \emph{graduated} structures, in which the rate rises in steps as the base grows larger. 

Second, the rate itself depends on the policy objective (cf. \Cref{sec:case_for_AI}). 
When the goal is corrective, policymakers often try to set the rate close to the marginal external cost of the harm, though that cost is often unknown, varies across contexts, or is hard to condition on all relevant factors~\cite{pigou1920economics_welfare, baumol1988theory, fleischer2015curb}. 
A uniform proxy rate may then serve as a second-best approximation (e.g., when the exact social cost varies by location). 
When the goal is instead redistributive or capacity-raising, the rate need not track marginal harm as closely. 
It may be set to capture windfall returns from AI or raise enough revenue to fund oversight,
though the rate should be set low enough to minimally achieve the goals and avoid unnecessary market distortion. 

Third, policymakers can choose how the schedule is phased and scoped over time and place via, for instance, gradual phase-ins, caps on total liability, entry thresholds below which no tax is owed, and location-specific rates that vary with local conditions. These choices add flexibility to tax rate design by adding ``discrete'' choices (whether to relocate, restructure, or stay just below a cutoff).

\textbf{Exemptions and credits.} Exemptions and credits 
determine which actors or activities are spared, either to avoid unintended burdens or to preserve beneficial uses.
For example, a tax aimed at labor-displacing AI adoption may choose to distinguish augmentation from substitution;
or a tax targeting profit-driven AI externalities may choose to exempt open-weight AI models. 
Exemptions can provide explicit carveouts to avoid burdening specific actors or activities,
and credits can incentivize certain behaviors not captured by the tax base. 
While actors can also reduce the measurable activity on which the tax is levied, 
exemptions and credits capture dimensions that are either less measurable or less directly tied to the harmful activity.
In this way, they are important design parameters that make taxation more flexible and precise.

\textbf{Revenue use.} Taxes may flow to general revenue or be earmarked for specific purposes. 
The specific purpose could be AI-related, 
such as \emph{funding programs related to worker transition; compensating creators; improving electricity and water infrastructure; and supporting capacity for AI regulations}. 
Alternatively, the earmarked purpose could be unrelated to AI, such as \emph{funding public education or welfare programs}. 
Unless a statute earmarks revenue, funds may not be directed as intended, but not earmarking it also allow tax revenue to be allocated flexibly based on changing needs.

\subsection{Mapping externalities to tax bases and instruments}\label{ssec:ai_instruments}

The parameters in \Cref{ssec:design_choices} combine differently for each externality. 
Below, we map externalities to plausible tax bases and instruments. 
We work through labor displacement and data-center resource burdens in detail, comparing how payroll, consumption, and excise taxes attach at different points in the AI supply chain~\cite{hopkins2025ai, cen2025mapping}; we then briefly discuss additional settings where rent taxes, excess-profit levies, benchmark-based exemptions, or non-tax responses may be more appropriate.

\subsubsection{Possible tax designs targeted at labor displacement's effect on payroll taxes}\label{ssec:tax_design_ex_1_labor}

As discussed in \Cref{ssec:labor_externality},
there are multiple externalities associated with labor displacement. 
Of these, we focus briefly on the third mentioned in \Cref{ssec:labor_externality}: that replacing human workers with AI reduces payroll taxes that corporations pay and the corresponding government revenue. 

Specifically, payroll taxes include contributions from both employers and employees, 
and the tax obligation is based \emph{only on the amount paid to human employees}. 
As a result, employers \emph{reduce their tax obligation} by replacing human employees with AI~\cite{bearer2025sharing}.
In other words, \emph{replacing human employees with AI reduces the amount of tax that employers have to pay}, on top of externalities caused by labor displacement. 
Moreover, the revenue from payroll taxes is generally allocated to programs that support workers, 
including retirement and healthcare, 
meaning that the reduction in payroll taxes due to AI displacing labor ironically coincides with a decrease in revenue for programs that could support workers.

Thus, adapting the payroll tax to require that employers are not incentivized to replace humans with AI or using an alternative instrument that captures this incentive gap is a natural fit. 
There are several considerations when choosing an instrument, including the existing \emph{infrastructure}, the ability to \emph{measure} the chosen tax base, and the intended \emph{purpose} of resulting tax revenue. For instance,
\begin{enumerate}
    \item  A payroll tax is the most conceptually natural fit, not only because it is the source of decreasing tax revenue, but also because payroll tax revenue fund social welfare programs for human workers (such as retirement and healthcare), which are programs that may become even more strained with labor displacement.
    However, the choice of tax base and corresponding measurement is less straightforward, 
    as it may be difficult to measure the ``pay'' of an AI ``worker''. 

    \item While the payroll tax may be the most conceptually natural fit (as discussed above), 
    ``payroll'' for AI may be dificult to measure, suggesting that more measurable quantitites, such as profits would be a better fit. 
    Corporate income, excess-profit, windfall-profit, or rent taxes are all candidates based on profits. 
    A general corporate income tax risks taxing all profits equally (both AI-related and non AI-related profits), especially for firms with diversified business models. 
    Thus, alternative approach could consist of taxing AI-related gains above a threshold~\cite{korinek2026publicfinance}. 
    For example, excess-profit taxes capture profits above a ``normal'' level, 
    and 
    windfall-profit taxes capture profits due to unusual gains.
    Taxing AI-related gains above a predetermined threshold could be done, e.g., through AI attributable profits reporting or by requiring AI activities to be organized in a dedicated subsidiary~\cite{bearer2025sharing}.

    \item However, taxing profits does not directly capture the intended activity, 
    which is replacing human labor with AI.
    One may argue that taxing profits unnecessarily penalizes companies from making good business decisions (\emph{for this particular taxation context of labor displacement}).
    An alternative is a \textbf{consumption tax} on corporations' use of AI, as this use is what causes labor displacement.
    For instance, one could place a consumption tax on API usage, which could be measured via tokens (per-unit tax) or the monetary cost of usage (ad valorem tax).

    Yet there are some important considerations. 
    For example, placing the consumption tax on API usage alone prices use of external models and could, consequently, encourage internal AI development. 
    This could be addressed by complementing a consumption tax with an excise tax on data centers (see \Cref{ssec:excise_data_center}). 
    Or one may believe that, as long as the consumption tax is appropriately low, companies for which AI is a tool (and not a core product) are unlikely to develop in-house AI from scratch because doing so is costly.
    Further discussion on consumption taxes can be found in \Cref{ssec:excise_data_center}.

    In addition to replacing lost payroll-tax revenue as discussed at the top of this section, a consumption tax would put a (small) price on replacing human labor with AI, incentivizing companies to more carefully consider when to use AI.
    Outside of the U.S., governments already treat digital and cloud-based services as taxable final consumption and collect the tax via the seller at the checkout~\cite{hmrc2022digitalservices}.
    To avoid an overly burdensome consumption tax, policymakers must carefully choose the rate, exemptions, and credits remains salient.
\end{enumerate}

\subsubsection{Possible tax designs targeted at data center externalities}\label{ssec:excise_data_center}

As discussed in \Cref{ssec:example_externalities_utilities}, 
AI's rising demand for compute creates multiple resource-related externalities. 
Data centers draw heavily on electricity and water, 
which can raise prices for nearby residents, 
strain local grids and water systems, 
and impose broader environmental costs that are not fully reflected in market prices. 
These harms scale with AI activity, 
which makes taxation a natural response.

Because these externalities scale with AI development, deployment, and use, it makes sense to tax in a way related to demand for natural resources.
There are several considerations when designing an appropriate tax, 
including the existing infrastructure, the ability to measure the chosen tax base, and where in the supply chain~\cite{hopkins2025ai, cen2025mapping} the tax attaches. 
For instance,
\begin{enumerate}
    \item 
    Because demand rises with use, one starting point is a \textbf{consumption tax} on AI services, such as a tax on token usage or broader API usage. 
    That is, for every token or API call, a tax is levied on the user (though the tax could be remitted by the model or API provider).
    This approach is effective at pricing downstream AI use. 
    However, such a token or API tax primarily touches \emph{inference} and \emph{downstream AI development}, 
    but it \emph{misses upstream training and development}.
    In other words, it would miss much of the AI development that produces models, such as OpenAI's GPT and Anthropic's Claude.
    Because much of AI model development occurs upstream,
    this approach does not provide full coverage. 
    It can also unintentionally encourage strategization among firms to avoid taxation, e.g., by redirecting development to other forms of compute that bypass APIs or avoid token usage.

    \item A second option is to move the consumption tax \emph{upstream}, to the purchase of data-center compute. 
    This version of a consumption tax is distinct from the token or API tax above;
    a data-center consumption tax is levied on the purchase of data-center compute. 
    That is, a firm would pay a tax for every purchase of data-center compute (e.g., server time or cloud compute), though the tax could be remitted by the data-center operator.
    This tax could be per-unit (e.g., based on the amount of server time rented) 
    or ad valorem (e.g., based on the total cost of the data-center compute).
    An ad valorem tax may be more natural, as tax liability would scale with how much compute a firm buys and the value of that compute (i.e., unlike a per-hour tax, a ``sales tax'' on data-center compute purchases would reflect the supply and demand for that compute, as not all compute may be equally valuable).
    
    This version of a consumption tax targets data-center usage more directly, and it reaches both training and inference.
    Its main weakness is evasion through vertical integration in that large firms may build or operate their own data centers to avoid the tax;
    this outcome could be avoided by taxing any type of data-center use (even internal use),
    though this design would only make sense as a per-unit tax.

    \item A third option is an \textbf{excise tax} on electricity and water use at data centers~\cite{due1994excise, crs_r43189}. 
    The tax base \emph{could} target AI-related electricity and water use, but it need not do so.
    If the primary goal is to address data centers' demand for natural resources, then the tax base could include all data-center related activity.
    On the other hand, if the primary goal is to target AI-related externalities and to fund AI regulations, then the tax base should be limited to AI activity (though discerning AI from non-AI activity may be definitionally and administratively challenging).
    Excise taxes are often imposed \emph{per unit} (e.g., per kilowatt-hour or per gallon of water withdrawn) rather than ad valorem. 
    Further, the data-center operator would likely be both the taxpayer and tax remitter under this setup. 

    The disadvantage of this setup is that the tax burden is not directly borne by the AI developers, deployers, and users who generate the harmful activity. 
    The tax burden could ultimately be passed to AI developers, deployers, and users via higher input prices that data-center operators impose on their consumers. 
    This indirect incidence is a disadvantage if policymakers want liability to sit with a particular actor in the supply chain~\cite{hopkins2025ai, cen2025mapping}. 
    The advantage is coverage: a resource-based excise can reach \emph{all} electricity and water tied to AI operations, including training, inference, and on-site facilities, not only tokenized API calls or third-party cloud purchases.
\end{enumerate}
These designs differ along several of the parameters in \Cref{ssec:design_choices}. 
They use different bases---tokens, compute hours, or resource consumption---and assign liability to different taxpayers, 
including AI users, AI developers, and data center operators. 
They also attach at different points in the AI supply chain~\cite{hopkins2025ai, cen2025mapping} and rely on different instruments, 
primarily consumption taxes versus excises. 
Choosing among them therefore requires tradeoffs between measurability, coverage, and who ultimately bears the burden.

Determining the right rate is a separate design problem. 
For corrective purposes, 
the rate should reflect the marginal external cost of the harm being targeted, 
whether that is the burden on local ratepayers, 
the cost of grid or water infrastructure, 
or environmental damage. 
Those costs vary by location and are often difficult to estimate precisely.

Two additional cautions apply. 
First, policymakers may wish to avoid double taxing environmental harms that are already priced via ``green'' taxes, 
while still addressing the externalities discussed in \Cref{ssec:example_externalities_utilities}. 
Second, data-center taxes create incentives to relocate computation to lower-tax jurisdictions, 
whether another state or another country. 
Coordinated adoption across jurisdictions would reduce this leakage; 
in the absence of coordination, 
mobile firms may simply shift where they train and run models.
This motivates taxing consumption since that is less easily relocated, 
but consumption taxes face the same challenges discussed above. 
\emph{As a result, multiple complementary tax instruments may be appropriate in this case.}

\subsubsection{Additional settings}

There are several additional externalities that may be more difficult to price, 
such as privacy harms, the spread of misinformation, and others named in \Cref{ssec:other_potential_externalities}.
In addition to the discussion of payroll, consumption, and excise taxes discussed above, we review several additional externalities and instruments below. 

\textbf{Rent taxes to counter market concentration.}
Rent taxes are designed to capture unusually high returns tied to scarcity or market power rather than ordinary competitive profit~\cite{garnaut2010principles}. 
They may apply to certain AI firms because frontier training and deployment require scarce inputs---compute, energy, water, data, and talent---at a scale that smaller rivals cannot readily match. 
Heavy resource demand and limited access to those inputs can constrain entry and rivalry, so a fully efficient competitive market may not be achievable; a rent tax offers one way to recapture extraordinary gains from concentration.

\textbf{Excess-profit and windfall taxes on ``large model'' risks.}
Some of the greatest risks arise from large models with dual-use capabilities: the same systems that enable powerful beneficial applications can also enable serious harmful ones, and preventing or controlling every negative use is impossible. 
These include catastrophic, biological, and misinformation risks. 
Excess-profit and windfall taxes (which apply to profits that exceed some level of ``ordinary'' profit) may be appropriate here because they can target the large firms that capture the greatest returns from developing and deploying models at this scale. 
They do not, however, distinguish between positive and negative uses.
Because they apply to profits broadly rather than to specific harmful acts, they do not price particular externalities at the margin. 
They could still extract revenue from the firms that benefit most from large-scale development, including cases where very large models generate widespread, diffuse, and hard-to-trace harms that cannot be attributed to any single deployment. 
Although policymakers could target large developers by instead tying liability to AI systems that exceed certain size thresholds (e.g., training compute or parameter counts), such proxies are brittle and invite restructuring around the cutoff.

\textbf{Benchmark-based exemptions.}
As discussed above, AI taxes may need to be more targeted than a broad corporate income tax or excess-profit levy. 
One approach is to grant exemptions or reduced liability when firms show that their systems perform well on specified benchmarks---for example, safety, fairness, or privacy metrics---so that the tax falls more heavily on systems that fail to meet a public standard. 
That can make the instrument more precise than a blanket profit tax. 
But the design can be fragile because benchmarks and thresholds must be chosen carefully, updated over time, and monitored for gaming; otherwise, firms may satisfy the formal test while changing little about underlying risk.

\textbf{Externalities that may not be taxable.}
Some harms may be poor candidates for taxation at all. 
Privacy loss and discriminatory decision-making are examples where assigning a dollar value risks treating harm as a permissible cost of business up to a measured threshold, which firms can then weigh against expected gains in ordinary cost-benefit calculations. 
The objection is not only that these harms are hard to measure. 
Rather, for some rights-based harms, pricing may be wrong in principle because it suggests that a priced amount of violation is acceptable; for others, the loss to victims is so asymmetric relative to any feasible payment that a tax cannot approximate the harm without trivializing it.

\subsection{Persistent question: AI supply chain and defining AI actors and activities}

A persistent challenge in the designs above is defining who and what to tax. 
Because AI development and deployment is distributed across a long supply chain of actors~\cite{hopkins2025ai, cen2025mapping, widder2023dislocated, cobbe2023understanding, bommasani2024ecosystem}, the boundaries between different AI actors and activities are often blurry in practice.
Some actors curate data while others pre-train models, and yet more post-train and adapt models for various use cases and applicatios. 
This poses a problem for tax design, 
as defining the tax base and taxpayer often requries drawing hard lines. 

For instance, consider a tax intended to target large AI developers.
Defining the tax base and taxpayer appropriately is challenging.
For one, how should one define ``developer''? 
Should a firm that fine-tunes a pre-trained model be considered a developer? 
Does it matter if the pre-trained models is open-weight or closed-weight? 
What about methods that leverage sophisticated in-context learning or prompting techniques that do not require explicit training?
Should ``AI-ness'' be determined by the architecture (e.g., presence of transformers)?
If one wishes to distinguish by size, should it be based on the size of the model, the amount of data used, the amount of training or compute applied, the revenue of the firm, or some other metric?
Most definitions and thresholds around AI are brittle. 
In practice, they risk creating boundaries that will be gamed by firms to avoid tax liability or will become quickly outdated.
Firms can relabel systems or make minor design changes to sit just outside the boundary while preserving much of the same functionality.

Similar problems arise in distinguishing AI from non-AI activities. 
For instance, if firms are taxed based on their AI-related activities, what counts as an AI-related activity?
Alternatively, if firms are taxed because they are an AI firm, should all activities that they perform as an AI firm be taxed?
Decisions around these questions can have unintended consequences as well.
If firms are taxed based on AI-related activities, 
then they may rebrand as many internal activities as possible as non-AI.
Alternatively, if firms are taxed for being AI firms, then they may be incentivized to restructure and separate into multiple specialized firms (e.g., vertical ``disintegration'').
In some cases, these outcomes are not negative; careful tax design requires considering these outcomes and determining whether they are acceptable.

One option is to define the tax base by observable economic quantities instead of technical categories. 
Excess-profit or windfall taxes, broad consumption taxes on AI usage, and excise taxes on data center electricity and water use do not require policymakers to resolve every dispute about training method, model type, or size.
These proxies are easier to administer, but they can tax activity that is not closely tied to the targeted harm and miss harmful activity that does not pass through the chosen base.

A second option is to scope liability based on firms' own market claims. 
For instance, instead of opting for policymakers to create boundaries or definitions around ``AI-ness'', 
one could categorize firms or activities based on a firm's market claims. 
Since many firms are incentivized to market themselves as producing AI models, data, or services, 
this approach may (i) be easier to administer, and (ii) more closely align policymaking objectives with firms' financial objectives. 
For example, if a firm markets itself as selling an AI product, 
then all of its data-center activities could be taxed. While some may view this as too broad, 
it avoids issues that arise when delineating which data-center activities count as ``AI-related'', and it disincentivizes firms from overclaiming their AI qualities.

\subsection{Discussion}

The tax instruments discussed above should not be read as mutually exclusive alternatives from which policymakers must select one. Rather, they identify different points at which the AI economy could be taxed. A corporate surcharge or rent tax targets the gains from AI, a consumption tax targets the purchase or use of AI services, and an excise tax or user fee targets a more specific activity, input, or proxy associated with harm. 
Thus, one might prioritize first determining which observable tax base most closely corresponds to both (i) the externality being addressed and (ii) to the policy objective being pursued.

In other words, the same externality may be addressed through different instruments for different reasons, 
meaning the choice of instrument should match the intended policy objective. 
A tax aimed at discouraging harmful conduct should be tied as closely as possible to the activity generating the harm. A tax aimed at compensation or redistribution may instead focus on capturing part of the gains generated by AI, even if it does not directly change firm behavior. A tax aimed at regulatory capacity may operate more like a user fee or safety levy, raising revenue from the firms whose activities create new oversight needs. 
Table~\ref{tab:matching_externalities_tax_instruments} therefore does not recommend a single tax design. It illustrates how different harms could be paired with plausible tax bases, instruments, and design constraints.

It is worth noting that these different tax mechanisms target diverse parts of the economy. The practical distinction and trade-offs between this collection of tax instruments becomes clearer when considering a concrete externality or situation.
For example, labor displacement illustrates why the choice of tax instrument depends on the intended function of the tax. A consumption tax on AI services would operate on the demand side by raising the cost of AI adoption, but it could also affect beneficial forms of augmentation. An excise tax could in principle be more targeted if tied to observable proxies for labor-substituting deployment, though such proxies would be difficult to define. A corporate surcharge or rent tax would influence automation decisions less directly, but could capture part of the gains from AI-enabled productivity and redirect them toward worker transition, unemployment insurance, or reskilling. The same logic applies to other AI externalities. 
The closer the tax base is to the harmful activity, the stronger the corrective rationale; the closer it is to profits or revenues, the stronger the redistributive or regulatory-capacity rationale.

We do not take a position on which should be preferred by policymakers. The objective of this section is to provide some clarity on the variety of available tax instruments and their related considerations. Because of varying institutional priorities and administrative capacities, further analyses beyond the scope of this paper must be performed, to select the appropriate policy tools for each harm.

%% file: sections/pitfalls_open_questions.tex
\section{Potential Pitfalls and Open Questions}\label{sec:pitfalls}

Despite the significant advantages of taxation, important concerns remain regarding feasibility, implementation, and potential unintended effects. These limitations and open questions must be acknowledged explicitly.

To begin with, taxation and government intervention can, when poorly designed, reduce incentives for investment, innovation and economic growth. This concern is especially important in the context of AI, since the technology is expected to contribute to the long-term productivity growth and GDP expansion~\cite{filippucci2024impact, aghion2017artificial, oecd2025adoption}. However, a tax on AI could increase the cost of adopting or developing AI systems, thereby slowing capital investments and productivity, affecting the broader economy~\cite{vartia2008taxes, stantcheva2021effects, hanappi2023does, bonaparte2026tax_robots}. For this reason, \textbf{an AI tax risks weakening one of the key potential drivers of future economic progress}, possibly obstructing some of the contemporary goals and priorities that depend on it, such as the improvement of living standards~\cite{oecd2026practical_guide_investment_tax_incentives, korinek2026publicfinance}.

Moreover, AI taxation may also weaken national competitiveness. This can occur through two channels: first, if firms remain in the taxing jurisdiction, higher development and deployment costs may reduce investment in R\&D, slow innovation, and ultimately weaken position relative to foreign competitors~\cite{millot2020corporate}. Second, if AI firms can relocate mobile activities, taxation may encourage ``jurisdictional shopping'' or leakage. In other words, firms could shift their investments, infrastructure (e.g., data centers, model training, IP ownership, etc.), or reported profits to lower-tax jurisdictions~\cite{bonaparte2026tax_robots}. In either scenario,\textbf{the domestic economy may capture smaller productivity and innovation gains from AI innovation, harming its competitive position on the global stage}. This risk is especially salient for location-specific tax instruments, such as corporate income taxes or taxes tied to domestic AI infrastructure. This may reinforce the previous concerns around poorly designed taxation that could slow economic growth\cite{bonaparte2026tax_robots}. 

Additionally, a feasibility concern is that the political process may shape an AI tax in favor of the firms most affected by it, weakening its original corrective or revenue-raising purpose. AI firms are often well-resourced and technically sophisticated, while the intended beneficiaries of redistributed tax revenues---such as workers, consumers, taxpayers, and affected communities---are diffuse~\cite{olson1971logic}. Public-choice theory and ``capture theory''~\cite{stigler2003theory, peltzman1976toward, dal2006regulatory} suggest that \textbf{firms have the incentives, capacity and power to shape tax design in their favor} (e.g., through narrow tax bases, exemptions, credits, or reporting proxies)~\cite{richter2009lobbying}. The risk is that an AI tax may be enacted only in a weakened form, protecting incumbents, shifting costs to smaller deployers or users, and undermining its corrective or redistributive purpose. Evidence that corporate lobbying is associated with lower effective tax rates reinforces this concern~\cite{richter2009lobbying}.

Even when policymakers resist political influence, taxation may still introduce distortions when design and implementation are poorly calibrated~\cite{oecd2026practical_guide_investment_tax_incentives}. \textbf{Three mechanisms may cause distortions: definitional problems, discontinuous behavioral responses, and unintended incidence effects}. First, an unclear tax base might make the tax either overinclusive or underinclusive. As critics of robot taxes highlight, even the term ``robot'' is difficult to define~\cite{mann2019considering}; this problem is even greater for AI, since AI systems are intangible, embedded across many applications and geographically distributed. A broad definition of ``AI system'' could tax activities unrelated to the targeted harm, while a narrow definition could invite avoidance. Second, taxes may produce non-linear responses when firms face kinks or threshold decisions~\cite{saez2010taxpayers, kleven2016bunching, devereux1998taxes}. Therefore, small tax changes can create large behavioral adjustments (e.g., relocation, restructuring or reclassification) instead of proportional reductions in the harmful activity. Finally, poor design can create unintended incidence effects: depending on elasticities and market structure, the real burden may fall on actors other than those formally liable for the tax~\cite{seligman1927shifting_incidence_taxation, fullerton2002tax, auerbach2006bears}. In the AI context, this could become regressive if the cost is shifted from firms onto lower-income users or smaller downstream deployers, worsening existing inequalities.

Furthermore, \textbf{a separate tax implementation problem concerns measurement}. Generally, a tax's effectiveness depends on the state's ability to identify the taxable activity, quantify it, and link it to the relevant harm. This is particularly important for corrective taxation: in principle, a well-designed Pigouvian tax tracks the marginal external cost of the activity being taxed~\cite{pigou1920economics_welfare, baumol1988theory, heine2012environmental}. However, some AI-related externalities, such as environmental costs, may appear easier to measure than others, such as misinformation. Yet even these inputs require precise measurement infrastructure (e.g., metering systems, reporting standards, etc.)~\cite{organization341measuring} for attributing resource use to AI rather than to other data-center activities. Moreover, the same physical input can generate different social costs across contexts (e.g., one gallon of water in a water-stressed region does not impose the same harm as in a region where water is more abundant)~\cite{li2025ai_water_footprint, mansur2012value, ebert2026ai}. As a result, a uniform excise tax based on proxies may misprice the externality and cause further misallocation~\cite{fleischer2015curb, fullerton2008environmental}. Although this concern is most acute for Pigouvian taxes, measurement problems can also extend to other tax instruments.

A final limitation consists of the boundaries of price-based governance. While taxation assigns a cost to harmful conduct, it may not require firms to change the underlying practices that generate the harm. As a result, \textbf{firms may treat the tax as a price for doing business and therefore as a moral license, rather than a signal to change the harmful practice}~\cite{gneezy2000fine, parker2006compliance}. This concern is especially salient where the harms are irreversible even when compensated, such as severe natural resource depletion, labor market displacements, or unsafe AI development. Moreover, instead of inducing the intended behavioral change, taxation may instead incentivize firms to invest efforts in tax planning, optimization or restructuring to minimize liability~\cite{slemrod2002tax}. 

Despite the limitations discussed above, this paper argues that these concerns do not rule out AI taxation altogether. Rather, they show that any AI tax would need to be carefully designed, targeted, and iteratively adjusted. Therefore, under those conditions, AI taxation may still offer a potentially valuable mechanism for addressing selected AI externalities, particularly where existing regulatory tools struggle to internalize costs, fund oversight, or support affected communities. For greater efficiency, taxation could be complemented with additional policy instruments to address objectives it cannot target directly, and supported with international tax coordination~\cite{bonaparte2026tax_robots}.

%% file: sections/conclusion.tex
\section{Conclusion}

Within the broader landscape of AI oversight, AI taxation offers several regulatory advantages: it can rely on existing administrative, audit, enforcement infrastructure; it is a highly flexible instrument, allowing policymakers to adjust design choices based on goals; and it can generate revenue to support complementary regulatory efforts. This paper shows that AI taxes need not take a a single form, nor must they be Pigouvian in design. Rather, tax instruments can incorporate Pigouvian features, while also serving broader fiscal, redistributive, and governance objectives. On balance, we argue that AI taxation deserves serious attention from policymakers as one component of a wider regulatory toolkit for addressing AI-related externalities.
The strongest case is for targeted and carefully designed instruments aimed at specific harms. In this sense, AI taxation should be understood not as a fixed proposal, but as a set of instruments, including excises, consumption and corporate/rent taxes, that may address different tax bases and serve different governance functions depending on the externality at issue. At the same time, AI taxation remains an imperfect governance tool: challenges include measurement difficulties, incidence effects, leakage, political capture, and potential costs to innovation and competitiveness. Nevertheless, this paper argues that where AI systems generate social costs that are not internalized by firms or addressed by existing regulation, carefully designed tax instruments may provide a useful complementary mechanism. Therefore, the case for AI taxation is not that taxes can govern AI alone, but that they can help reveal, price, and potentially correct some AI externalities while funding the public capacity needed to manage them.

%% file: sections/acknowledgements.tex
\section*{Acknowledgments}

We thank Reed Schuler, Elham Tabassi, Serena Booth, and Senator Warren's office (in particular, Jenny Blessing, Daniel Ki, and Vivian Tarbert) for helpful comments and discussions.